\documentclass[12pt,preprint]{aastex}

\def\apj{{\rm ApJ}}

\def\mnras{{\rm MNRAS}}
\def\apjs{{\rm ApJS}}
\def\aap{{\rm A\&A}}
\def\araa{{\rm ARA\&A}}

\def\apjl{{\rm ApJL}}
\def\aj{{\rm AJ}}
\newcommand{\xmm}{\emph{XMM-Newton}}
\newcommand{\cha}{\emph{Chandra}}
\usepackage{graphicx}
\usepackage{longtable,lscape}
\usepackage{amssymb}

\shorttitle{Fitting LINER nuclei within the AGN family: A matter of obscuration?}
\shortauthors{O. Gonz\'alez-Mart\'in et al.}

\begin{document}
\title{Fitting LINER nuclei within the AGN family: \\ A matter of obscuration?}
\author{O. Gonz\'alez-Mart\'in\altaffilmark{1}}
\affil{Physics Department, University of Crete, PO Box 2208, 71003 Heraklion, Crete, Greece \newline 
X-ray Astronomy Group, Department of Physics and Astronomy, Leicester University, Leicester LE1 7RH, UK}
\email{omaira@physics.uoc.gr}

\author{J. Masegosa and I. M\'arquez}
\affil{Instituto de Astrof\'isica de Andaluc\'ia (CSIC), Granada, SPAIN}

\and

\author{M. Guainazzi}
\affil{European Space Astronomy Centre of ESA, P.O. Box 78, Villanueva de la
Canada, E-28691 Madrid, SPAIN}

\altaffiltext{1}{Instituto de Astrof\'isica de Andaluc\'ia (CSIC), Granada, SPAIN}


\begin{abstract}
In this paper we study the nuclear obscuration of galaxies
hosting {\it Low Ionization Narrow Emission Regions} (LINERs) based on their
X-ray and optical emission. They show column densities at soft energies (0.5-2 keV)
mostly related to the diffuse emission around the AGN, showing a
correlation with the optical extinction. Column densities at hard energies
(2-10 keV) seem to be much higher than what would be expected from the optical extinction. They might
be associated to the inner regions of the AGN, buried at optical wavelengths. 
The main result of this paper is that around 
50\% of our LINER sample shows
signatures of {\it Compton-thickness}  according to the most common tracers: the
X-ray spectral index, $\sf{F_{X}(2-10~keV)/F([OIII])}$ ratio and $\sf{FeK\alpha}$ equivalent width 
(EW). However, the EWs of {\it Compton-thick} LINERs are 
significantly lower than in {\it Compton-thick} Seyferts ($\simeq$200~eV against $\ge 500$~eV), 
suggesting that the 2--10~keV emission is dominated by electron scattering of the 
otherwise invisible AGN, or by emission from shocked gas associated to star 
formation rather than by reflection from the inner wall of the torus.
However, no clear relation seems to exist between galaxies with optical dust lanes
and X-ray classified {\it Compton-thick} objects. This may suggest that \emph{Compton-thick} sources 
should be related to absorbing material located at the very inner 
regions of the AGN, maybe in the putative dusty torus.
Larger black hole masses and lower Eddington ratios than Seyfert
galaxies have been found. This effect can be better attributed to LINER nuclei being hosted by earlier morphological
types than Seyfert nuclei. However, it has to be noted
that, once a proper correction to the X-ray luminosity is applied, LINERs show Eddington ratios 
overlapping those of type 2 Seyferts.
We speculate with a possible scenario for LINER nuclei:
an inner obscuring matter 
similar to that of type 2 Seyfert, and an external obscuring matter responsible 
for the optical extinction. {\it Compton-thick} sources
appear to be more common among LINERs than Seyferts.
\end{abstract}

\keywords{LINERs -- AGN -- X-rays -- \cha{} -- \xmm{}.}

\section{Introduction}

Active Galactic Nuclei (AGN) emit over the entire electromagnetic
spectrum and are widely believed to be powered by the accretion of matter
onto a super-massive black hole \citep[SMBH,][]{rees_black_1984}. Several families within
the AGN category have been established from an observational point of
view. Although their classification is sometimes misleading, it is
widely believed that an unified model can explain
them under a single scenario \citep{antonucci_unied_1993}. A key ingredient in this
scheme is a dusty torus whose inclination with respect to the
observer's line of sight is responsible for the dichotomy between optical type 1 (with broad
permitted lines, face-on view) and type 2 (with narrow permitted
lines, edge-on view) AGN. However, this scheme needs to be further
confirmed because there are several sub-classes of objects that cannot be easily fitted
into it. As an example, the nature of Narrow-line Seyfert 1 \citep{dewangan_2005}
or non-obscured Seyfert 2 \citep{panessa_unabsorbed_2002} is still a matter of debate. 

{\it Low Ionization Nuclear
Emission Line Regions} (LINERs) are another sub-class of objects that cannot be 
easily included in the unified model. They are intriguing cases
because, as suggested by their low X-ray luminosities ($\rm{L(2-10~keV)}$ $\sf{\sim 10^{39-42}erg~s^{-1}}$),   
they could be the link between AGN ($\rm{L(2-10~keV)}$ $\sf{\sim 10^{41-45}erg~s^{-1}}$) 
and normal galaxies \citep{Zhang_2009,Rovilos_2009}.
Furthermore, they are the dominant population of 
active galaxies in the nearby universe \citep{ho_search_1997, ho_new_2008}. Their  
signature in the optical spectrum is 
the enhancement of low ionization lines. However, this property 
alone is not enough to disentangle the nature of these galaxies because it 
can be explained by a variety of different physical processes 
\citep[][]{ho_new_2008}. Compact radio \citep{nagar_radio_2005} and hard 
X-ray cores \citep[][and references therein, hereinafter GM+09]{gonzalez-martin_2009} 
are the most secure signatures for the presence of an AGN.
Although the AGN nature of a
large number of LINERs has been confirmed from data at X-ray
and radio frequencies, it is still unclear how
LINERs do fit into the AGN unified scenario.
A radiatively inefficient 
accretion flow onto the SMBH and/or a large amount of obscuring matter 
have been proposed as the main differences between LINERs 
and more luminous AGN \citep{Dudik_2008}. X-rays are the ideal laboratory to test their nature, 
since they provide  valuable information on both the obscuration and
accretion rates.

We have analyzed the largest sample of LINERs up to now
at X-ray frequencies (82 objects) with \cha{} and \xmm{} data 
(GM+09). {\it Chandra}'s excellent 
angular resolution allowed us to investigate the X-ray nuclear 
properties of these galaxies \citep[see also][hereinafter GM+06]{gonzalez-martin_x-ray_2006}. 
According to their nuclear morphology in the 4.5-8~keV band, we found 
that almost 60\% of the sample shows an unresolved nuclear source, which 
is a clear hint of their AGN nature. The addition of \xmm{} 
data offers us the opportunity to perform the spectral analysis on 60 out of the 
82 objects. From the X-ray point of view, we concluded that LINERs are similar 
to type 2 Seyferts, both in luminosity and spectral shape (GM+09). 

In this paper we discuss the properties and nature of the 
obscuring material covering LINER nuclei. In Section 2 we review the sample selection 
(already presented in GM+09); Section 3  presents the
observational tracers of {\it Compton-thick} obscuration for our sample; in 
Section 4 we correlate the X-ray obscuration with 
other multiwavelength observables, to derive further clues on the origin of the obscuring 
material; and in Section 5 we discuss the implications of the obscuration on the LINER
spectral energy distribution (SED) of LINERs. A summary of our results and the 
conclusions are provided in Section 6.

\section{The sample}							    
									    
Our sample is presented in GM+09, where all the observational details	    
are exhaustively explained. We briefly describe here the main characteristics 	    
of the sample. 							    
									    
The sample was extracted from the multi-wavelength LINER catalogue	    
compiled by \citet{carrillo_multifrequency_1999} (hereinafter MCL).	    
The sample includes all the galaxies in MCL with 	    
available \cha{} data up to 2007-06-30 and \xmm{} data up to		    
2007-04-30. It includes 108 LINERs with \cha{} data and 	    
107 LINERs with \xmm{} data. Seventy six objects are present in 	    
both archives yielding to a total of 139 LINERs.				    
									    
LINER identifications were revised using \citet{veilleux_high-resolution_1987} diagnostic diagrams	    
to discard out Seyfert, HII and transition objects.	    
After optical re-identification we ended up with a final sample of	    
83 sources including 68 observed with \cha{} and 55 with \xmm{}.	    
Observations for one of	    
these objects showed strong pile-up effects leaving us with		    
a final sample of 82 objects. Forty LINERs are found in both datasets. 
The final sample mainly comprises objects from the Palomar Survey 
\citep{ho_search_1997}	and Luminous and Ultra-luminous IR galaxies 
\citep[LIRGs and ULIRGs, mainly from ][ and reference therein]{veilleux_new_1999}. 
Note that this sample is not complete because it comes from a catalogue containing all the
known LINERs until 1999 and include only available data in \cha{} and \xmm{} archives.
									    
See GM+09 for further details on the X-ray observations (its Sections	    
2 and 3 and Table 2), previously published X-ray data     
(its Appendix B), spectral fits used      
along this text (its Sections 4.2, 4.3 and 5.1.2),    
F-test statistic (its Tables 3 and 4), an example of spectral fit (its Fig. 4), spectral  
fit figures for {\it Chandra} and {\it XMM-Newton} data      
(in on-line format in its Appendix D and E), and  final spectral fits (its Table 7).

\section{Compton-thickness}

The X-ray spectrum of LINERs can be described by two main 
components: {\it (i)} an absorbed primary 
power-law continuum; {\it (ii)} a soft spectrum 
(below 2 keV) described by an absorbed scattering plus/or a thermal component 
(GM+09). In this scenario, the column densities, 
called NH1 and NH2, provide information on the amount of 
absorbing material associated to the soft (0.5-2 keV) energy band and
to the hard (2-10 keV) energy band, respectively.

If the X-ray obscuring matter has a column density which is equal 
to or larger than the inverse of the Thomson cross-section 
($\sf{N_{H} \gtrsim 1.5\times10^{24} cm^{-2}}$), then the source is 
called, by definition, {\it Compton-thick}. If the X-ray obscuring matter has a
column density lower than the {\it Compton-thick} limit but still 
in excess to the Galactic one the source is called {\it Compton-thin}.
In {\it Compton-thick} AGN, the reflection components
can be misinterpreted as a primary continuum and consequently induce a
misclassification as {\it Compton-thin} or even unobscured AGN.
Since the intrinsic continuum in {\it Compton-thick} sources is detectable at 
energies $>$10 keV, only indirect proofs of the {\it Compton-thick} nature 
can be obtained with \cha{} and \xmm{} observations.

From our data analysis, we found that NH2 covers a wide range of values 
($\rm{log(NH2)=}$ $\rm{20-24~cm^{-2}}$), the range covered by NH1 is much narrower, 
with a median value of 
$\rm{log(NH1)}$ = 21.32$\rm{\pm}$ 0.71 (GM+09).  Both column densities, NH1 
and NH2, are given in Table \ref{cha3:tab1} (Cols. 4 and 5).
The reason for such a
behavior might be related to the nature and location of the
obscuration, as we discuss later.  
However, LINER nuclei can be {\it Compton-thick} sources and, in this case, 
the interpretation of the measured column density could be different.

The evaluation of {\it Compton-thickness} will be done based on three indirect diagnostics: 
(1) spectral index ($\rm{\Gamma < 1}$), 
(2)$\sf{F_{X}(2-10~keV)/F([OIII])}$ ratio ($\sf{log(F_{X}(2-10~keV)/F([OIII]))<0.5}$) and 
(3) high equivalent width of the neutral iron emission line 
   (EW(FeK$\sf{ \alpha) > 500 eV}$).
The relevant information of these three diagnostics and the 
classification according to them are reported in Table \ref{cha3:tab2}.

 \subsection{Spectral index}

A flat spectrum above $\sim$ 2 keV is one of the known {\it
Compton-thick} diagnostics \citep{maiolino_heavy_1998,cappi_x-ray_2006}. 
In addition to the models used in GM+09, in this paper we have fitted 
all the data to an absorbed power-law at energies larger than 2 keV to 
have an independent measurement of the possible flat spectrum.  Three 
Gaussian fits at 6.4, 6.7, and 6.95 keV have been included because 
FeK${\sf \alpha}$, FeXXVI and FeXXVII emission line are present in a 
number of objects (GM+09). The resulting spectral indices are listed 
in Table \ref{cha3:tab2} (Col. 2).  
It should be noticed that for NGC\,833, NGC\,835, NGC\,2639, UGC\,4881, NGC\,3507, NGC\,3898, 
NGC\,3945, MRK\,266NE and NGC\,6482 a spectral fit was reported in GM+09 
(more than 200 counts in 0.5-10 keV energy band),
but insufficient number counts above 2 keV prevents their analysis here. 
Thus, this analysis was possible in 51 out of the 82 LINERs.

We consider a spectrum as flat when the resulting spectral index is
consistent, within the uncertainties, with being smaller than 1.2 
\citep[][]{risaliti_2002b,Beckmann_2006,Dadina_2008,Winter_2009}.
Statistically, flat spectra have been identified for 20 out of the 51 
objects (39\%). 

\subsection{$\sf{F_{X}(2-10~keV)/F([OIII])}$ ratio}

\citet{maiolino_low-luminosity_1995} showed that the [OIII] line emission
can be considered as a good isotropic indicator of the AGN power.
On the other side, the 2-10 keV X-ray emission should be an intrinsic AGN
property in the cases where the primary continuum is not
suppressed by a highly obscuring material. When the primary
continuum is suppressed due to heavy absorption
($\sf{N_{H}>10^{24}cm^{-2}}$), the ratio between hard X-rays and [OIII] line emission
lowers because the computed X-ray luminosity is underestimated. For a large 
sample of type 2 Seyferts, the $\sf{log(F_{X}(2-10~keV)/F([OIII]))}$ ratio has 
been used as a good diagnostic to discriminate between
{\it Compton-thick} and {\it Compton-thin} sources 
\citep{bassani_three-dimensional_1999,panessa_unabsorbed_2002,maiolino_elusive_2003,
panessa_nature_2005,panessa_x-ray_2006}. We will use this criterion to search
for {\it Compton-thick} LINER nuclei.

We have searched in the literature for [OIII] emission line fluxes and
$\sf{H\alpha/H\beta}$ ratios. Data are available for 79 out of the 82 objects
(except for NGC\,835, CGCG\,162-010 and IC\,1459).  The optical extinction is 
computed using Av = $\sf{6.67\times log(f(H\alpha)/Rv\times f(H\beta)}$). 
Whenever the Balmer Decrement is lower than 3.1, we assume a value of 3.1, 
corresponding to zero optical extinction (Osterbrock 1987).
L(2-10~keV) is taken from GM+09. Reddening-corrected [OIII] emission line fluxes 
and optical extinctions are reported in Table \ref{cha3:tab1} (Cols. 7 and 8) and the 
$\sf{log(F_{X}(2-10~keV)/F([OIII]))}$ ratios are shown in Table \ref{cha3:tab2} (Col. 4).

\begin{figure}
\begin{center}
\includegraphics[width=0.9\columnwidth,height=0.9\columnwidth]{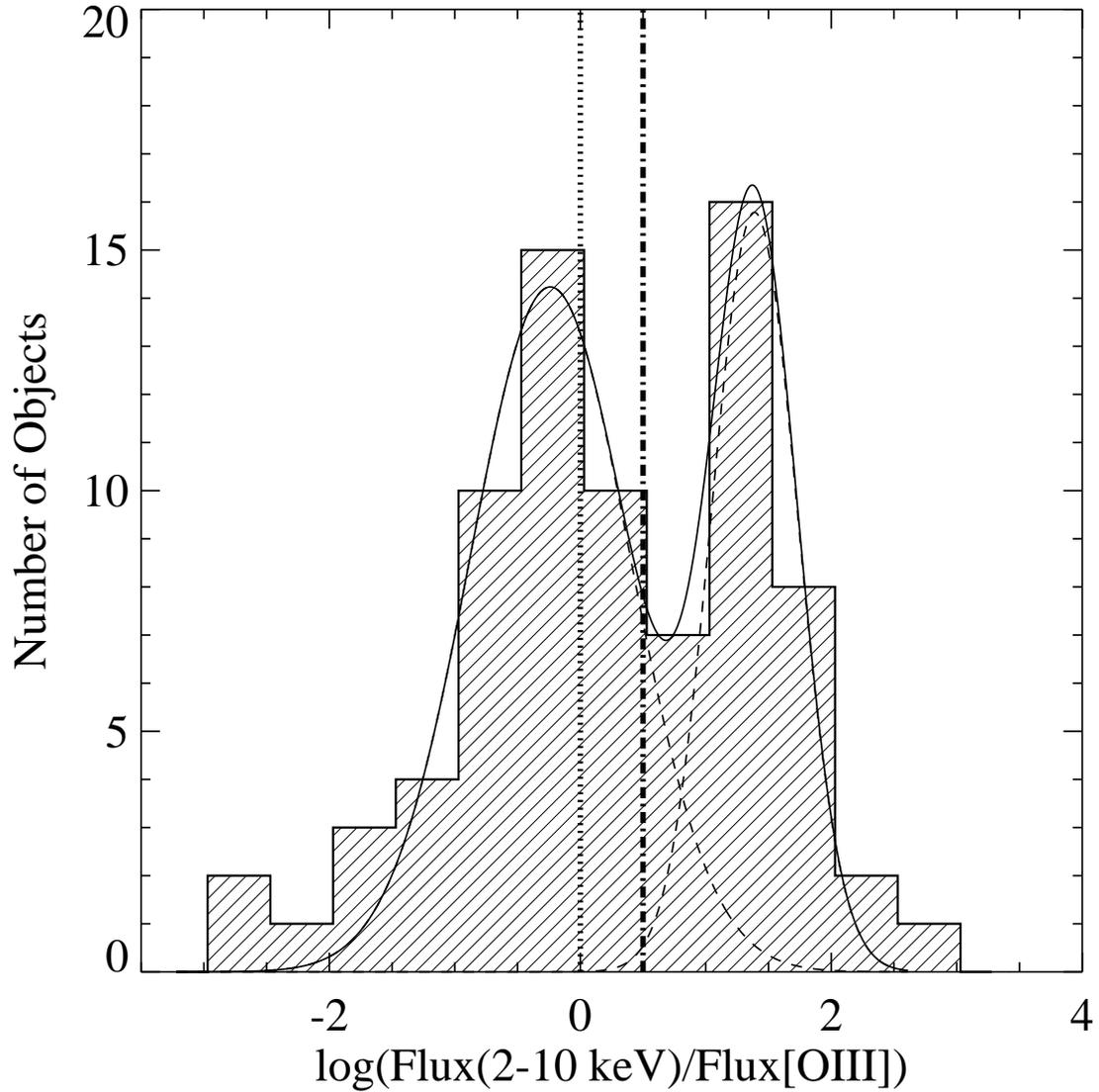}
\caption{Histogram for the ratio between X-ray 2-10~keV energy band fluxes and reddening-corrected
[OIII] emission line fluxes.  The dot-dashed line correspond to the limit found by \citet{maiolino_heavy_1998} between 
{\it Compton-thin} and {\it Compton-thick} sources. The dotted line correspond to the conservative 
limit adopted by \citet{cappi_x-ray_2006}. Continuous line shows the best fit to a two Gaussians model and the 
dashed lines show the Gaussian components. 
}\label{cha3:fig:histoLHOIII}
\end{center}
\end{figure}

Fig.~\ref{cha3:fig:histoLHOIII} shows the histogram of  
$\sf{log(F_{X}(2-10~keV)/F([OIII]))}$ ratios. 
We have fitted the distribution to a two Gaussian model 
(see continuous and dashed lines in Fig.~\ref{cha3:fig:histoLHOIII}). 
The double Gaussian model are centred
at $\sf{log(F_{X}(2-10~keV)/F([OIII]))_{o1}=-0.24}$ and $\sf{log(F_{X}(2-10~keV)}$ $\sf{/F([OIII]))_{o2}=1.39}$, with
$\rm{\sigma=0.6}$ and $\rm{\sigma=0.3}$, respectively. 
The minimum between the two Gaussian fits occurs 
at $\sf{log(F_{X}(2-10~keV)/F([OIII]))}$=0.68.
The distribution is
similar to that found by \citet{maiolino_heavy_1998} for 
Seyfert galaxies. It shows two peaks centered at $\sf{log(F_{X}(2-10~keV)/F([OIII]))=-0.5}$ 
and $\sf{log(F_{X}(2-10~keV)/F([OIII]))}$=1.0, with the later
value corresponding to the average ratio found for type 1 Seyfert galaxies
and the former to that of {\it Compton-thick} objects. 
Note that the Gaussian fit is centred at the {\it Compton-thick} regime 
($\sf{log(F_{X}(2-10~keV)/F([OIII]))_{o1}=-0.24}$) shows an 
area that represents the 63\% of the sample.

We have decided to define two regimes to ensure the robustness			   
of our classification: (1) $\sf{log(F_{X}(2-10~keV)/F([OIII]))<~0}$			   
and (2) $\sf{0}$ $\sf{<log(F_{X}(2-10~keV)/F([OIII]))<}$ $\sf{0.5}$. The value			   
$\sf{log(F_{X}(2-10~keV)/F([OIII]))}$ $\sf{<0.5}$ corresponds to the			   
limit reported by \citet{maiolino_heavy_1998} between type 1 Seyferts			   
and {\it Compton-thick} type 2 Seyferts. The ratio					   
$\sf{log(F_{X}}$ $\sf{(2-10~keV)/F([OIII]))<~0}$ is the conservative limit		   
assumed by \citet{bassani_three-dimensional_1999}.  
											   
\citet{maiolino_heavy_1998} presented a sample of 8 heavily absorbed type			   
2 Seyferts by means of {\it BeppoSAX} data. They measured their column			   
densities, spectral index and EW(FeK$\rm{\alpha}$), classifying all of them as			   
{\it Compton-thick} sources. They compared the ratio $\sf{log(F_{X}(2-10~keV)/F([OIII]))}$   		   
with that of type 1 Seyferts by \citet{Mulchaey_1994}. This ratio is 		   
$\sf{log(F_{X}(2-10~keV)/F([OIII]))}$ $\rm{\sim}$ 1 for type 1 Seyferts while {\it Compton-thick}  		   
sources always show $\sf{log(F_{X}(2-10~keV)/F([OIII]))}$ $\rm{<}$ 0.5		   
\citep[see Fig. 3 in ][]{maiolino_heavy_1998}. On the other hand, \citet{bassani_three-dimensional_1999}		   
compared the column density versus the ratio $\sf{log(F_{X}(2-10~keV)/F([OIII]))}$ for  72 type 2 Seyferts.  	   
They also refered to \citet{Mulchaey_1994} for type 1 Seyferts. They found		   
a ratio $\sf{log(F_{X}(2-10~keV)/F([OIII]))}$=0 for heavily absorbed sources. However, they 		   
compile their sample using all the type 2 Seyferts observed in the 2-10 keV		   
range. To establish the limit between {\it Compton-thick} and {\it Compton-thin} 	   
sources, we consider crucial that column densities of {\it Compton-thick}		   
sources must be confirmed above 10 keV. Thus, we have used the limit given by \citet{maiolino_heavy_1998}, since they used {\it BeppoSAX} data above 10 keV.

\begin{figure}
\begin{center}
\includegraphics[width=0.9\columnwidth,height=0.9\columnwidth]{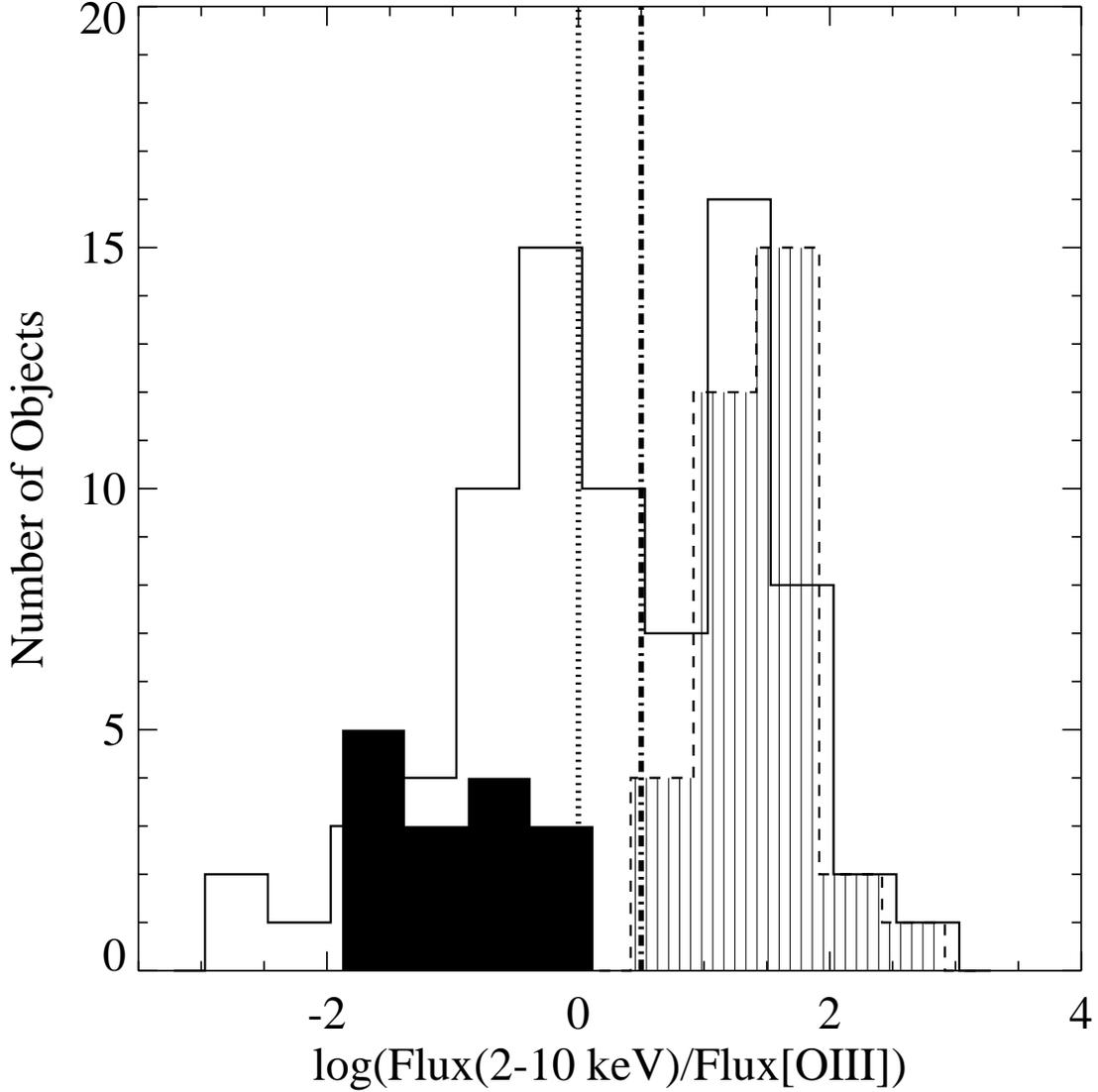}
\caption{Histogram for the ratio between X-ray 2-10~keV energy band fluxes and reddening-corrected
[OIII] emission line fluxes.  The dot-dashed line correspond to the limit found by \citet{maiolino_heavy_1998} between 
{\it Compton-thin} and {\it Compton-thick} sources. The dotted line correspond to the conservative 
limit adopted by \citet{cappi_x-ray_2006}. The empty histogram is the distribution obtained with our LINER sample. 
The black-filled histogram contains {\it Compton-thick} sources by Bassani et al. (1999) and the dashed filled 
histogram shows the unobscured PG QSOs reported by
\citet{jimenez-bailon_xmm-newton_2005} and \citet{piconcelli_2005}.}\label{cha3:fig:histoLHOIII_AGN}
\end{center}
\end{figure}

To validate such a limit we have compared the distribution of $\sf{log(F_{X}(2-10~keV)}$ $\sf{/F([OIII]))}$  
(see Fig. \ref{cha3:fig:histoLHOIII_AGN})
obtained for our LINER sample with that for unobscured PG QSOs \citep{jimenez-bailon_xmm-newton_2005,piconcelli_2005}
 and {\it Compton-thick} sources reported by \citep{bassani_three-dimensional_1999}. The [OIII] fluxes of PG 
QSOs are taken from \citet{Marziani_2003} and corrected for reddening assuming  $\rm{H\alpha/H\beta=4.5}$ 
\citep[mean value for QSOs by ][]{York_2006}. Unobscured QSOs show $\sf{log(F_{X}(2-10~keV)/F([OIII]))> 0.}$, in
agreement with our second
peak centred at $\sf{log(F_{X}(2-10~keV)}$ $\sf{/F([OIII]))_{o2}=1.39}$. {\it Compton-thick} AGN fall in the 
region of  {\it Compton-thick} LINERs. Moreover, \citet{Lamastra_2009} have recently shown 
that {\it Compton-thin} type 2
Seyferts range between $\sf{0<log(F_{X}(2-10~keV)/F([OIII]))< 3}$, overlapping with the PG-QSOs. 
											   
Thirty three out of the 79 LINER galaxies (42\%) have $\sf{log(F_{X}(2-10~keV)/F([OIII]))}$ $\sf{<~0}$  
following the more conservative criterion. Nine of them show ratios			   
$\sf{0}$ $\sf{<log(F_{X}(2-10~keV)}$ $\sf{/F([OIII]))<}$ $\sf{0.5}$. The fraction of LINERs with		   
$\sf{log(F_{X}(2-10~keV)}$ $\sf{/F([OIII]))}$ $\sf{<0.5}$ is then 53\% (42/79).

Two points must be stressed on the use of this diagnostic
to find {\it Compton-thick} objects. First, we notice that
21 sources have an unrealistic ratio of $\sf{H\alpha/H\beta}$ below
3.10, the assumed ratio for AGN \citep[][]{osterbrock_astrophysics_2006}. 
This might be due to 
a problematic continuum subtraction. Therefore 
in these 21 cases we have assumed a minimum value of 3.1,
which means that no correction has been performed. The [OIII] emission line
flux hence is underestimated, and therefore the number of {\it
Compton-thick} LINERs in our sample can be taken as a lower limit.

The second comment is about the eventual contamination by emission		 
from the host galaxy and/or circumnuclear environment. Objects		 
with a disrupted morphology, like merging galaxies, could have an		 
enhancement of star formation processes that over-shine the host 		 
galaxy emission. The usage of the [OIII] emission lines versus X-ray luminosity
ratio as a {\it Compton-thick} indicator could lead to misleading results.
A large fraction		 
of the measured [OIII] emission line flux could come from star-forming  	 
processes and the nuclear [OIII] emission line luminosity could be		 
overestimated.

LIRGS is well known that they show perturbed morphologies and an enhancement of star formation 
which can turn out in large values of [OIII] fluxes. Evidence of
perturbed morphologies is present in only 7 cases (IIIZW\,035,  		 
UGC\,4881, NGC\,3690B, IRAS\,12112+0305, NGC\,4410A, MRK\,266NE and		 
NGC\,6240, see GM+09\footnote{The post stamps of DSS images at 150 kpc  	 
scale are provided in Appendix F in GM+09.}), all of them classified		 
according to this criterion as {\it Compton-thick} candidates.  		 
NGC\,3690B and NGC\,6240 are well known {\it Compton-thick} sources	 
(see Section 3.1). The remaining 5 cases should be taken with reservation.

A contamination of the [OIII] flux measurement by merging-driven star formation should affect more 
significantly galaxies at a larger distance, because a larger fraction of the 
morphological disturbances are included in the spectroscopic aperture. However, the fraction of
under [OIII]-luminous LINERs in our sample does not depend on the source distance. Using objects with 
the smallest distance (D$\rm{<} 50 Mpc$) we get the same percentage: 27 out of		 
the 54 objects (50 per cent) show $\sf{log(F_{X}(2-10~keV)/F([OIII]))<0.5}$.
In fact, no trend is found between this ratio and the distance (correlation	 
coefficient r=0.15). Therefore, aperture effects can be ruled out.  
										 
Among unperturbed morphologies, evidences of circumnuclear star formation	 
have been found in only three cases: NGC\,3507, NGC\,3998 and NGC\,4321 	 
\citep{delgado_stellar_2004}. Thus, the {\it Compton-thick} nature for  	 
NGC\,3507\footnote{The only case among the three galaxies with  		 
$\sf{log(F_{X}(2-10~keV)/F([OIII]))}$ $\sf{<0.5}$)} could be questioned 	 
because an enhancement of the [OIII] line emission might be expected		 
due to the ionized emission in the central region. Recently,			 
\citet{Walsh_2008} discard such a possibility based on STIS/{\it HST} data:	 
the H$\alpha$+[NII] emission line morphology appears to be very compact			 
with no extended features attributable to circumnuclear star forming events.
										 
Furthermore, Ho (2008) remarked						 
that all the classes of LLAGN (Seyfert, LINERs and Transition objects)  	 
show the same host galaxy properties after a carefully decontamination 	 
of the differences coming from the Hubble type distribution of each		 
class. Only transition objects seem to show a mild enhancement of star  	 
formation of the host galaxy. This is also the case for their circumnuclear	 
environments. On nuclear scales smaller than $\rm{10 pc}$, \citet{Sarzi_2005} 		 
studied the stellar population of nearby LLAGN, finding that only 1 out 	 
of their 4 LINERs showed young stellar populations. \citet{delgado_stellar_2004,gonzalez-delgado_hst/wfpc2_2008}	 
 found also that LINERs host old stellar		 
population. Thus, recent star-formation does not seem to be an important ingredient	 
in LINERs. Therefore, in general terms we could expect that the [OIII] to	 
X-ray flux ratio would be as good tracer as already previous studies have shown 
for Seyfert nuclei (Maiolino et al. 1998; Cappi et al. 1999; Bassani et al. 2000;  	 
\citet{panessa_x-ray_2006}; Cappi et al. 2006). This result is reinforced because 
Seyfert 1, Seyfert 2 and unboscured
QSOs follow the same relation between 2-10 keV X-ray luminosity and [OIII] 
emission line luminosity than LINER nuclei (see Section \ref{sec:eddington}).

\subsection{EW of the neutral iron emission line}

\citet{Leahy_1993} found that $\sf{EW(FeK\alpha)}$ is
another {\it Compton-thick} tracer. The idea comes from the finding
that $\sf{EW(FeK\alpha)}$ for Seyfert 1 galaxies typically amount to
a few hundred eVs
\citep{turner_asca_1998,perola_compton_2002,panessa_broad-band_2008}.
When the column density increases to a few
$\sf{10^{23}~cm^{-2}}$, EWs increase, because they are measured against
a suppressed continuum. $\sf{EW(FeK\alpha)}$ can reach values as
high as 500~eV for column densities larger than $\sf{10^{24}~cm^{-2}}$
\citep{matt_x-ray_1997,bassani_three-dimensional_1999}.

For our sample of 82 LINERs, clear detections of FeK${\sf \alpha}$ lines are 
found in 10 galaxies with \xmm{} data and in
7 galaxies with \cha{} data, 4 of them in common \citep[see][]{gonzalez-martin_2008}. All together
we have positive detection for NGC\,0315, NGC\,0833,
NGC\,0835, NGC\,1052, UGC\,05101, NGC\,3690B, NGC\,4486, NGC\,4579,
MRK\,0266NE, UGC 08696, NGC\,6240, NGC\,7130 and NGC\,7285.
Their EW and uncertainties are presented in Table \ref{cha3:tab2}.
Only three out of the 13 detected FeK${\sf \alpha}$ line show 
EW${\sf >}$500 eV.

\begin{figure}
\begin{center}
\includegraphics[width=0.9\columnwidth,height=0.9\columnwidth]{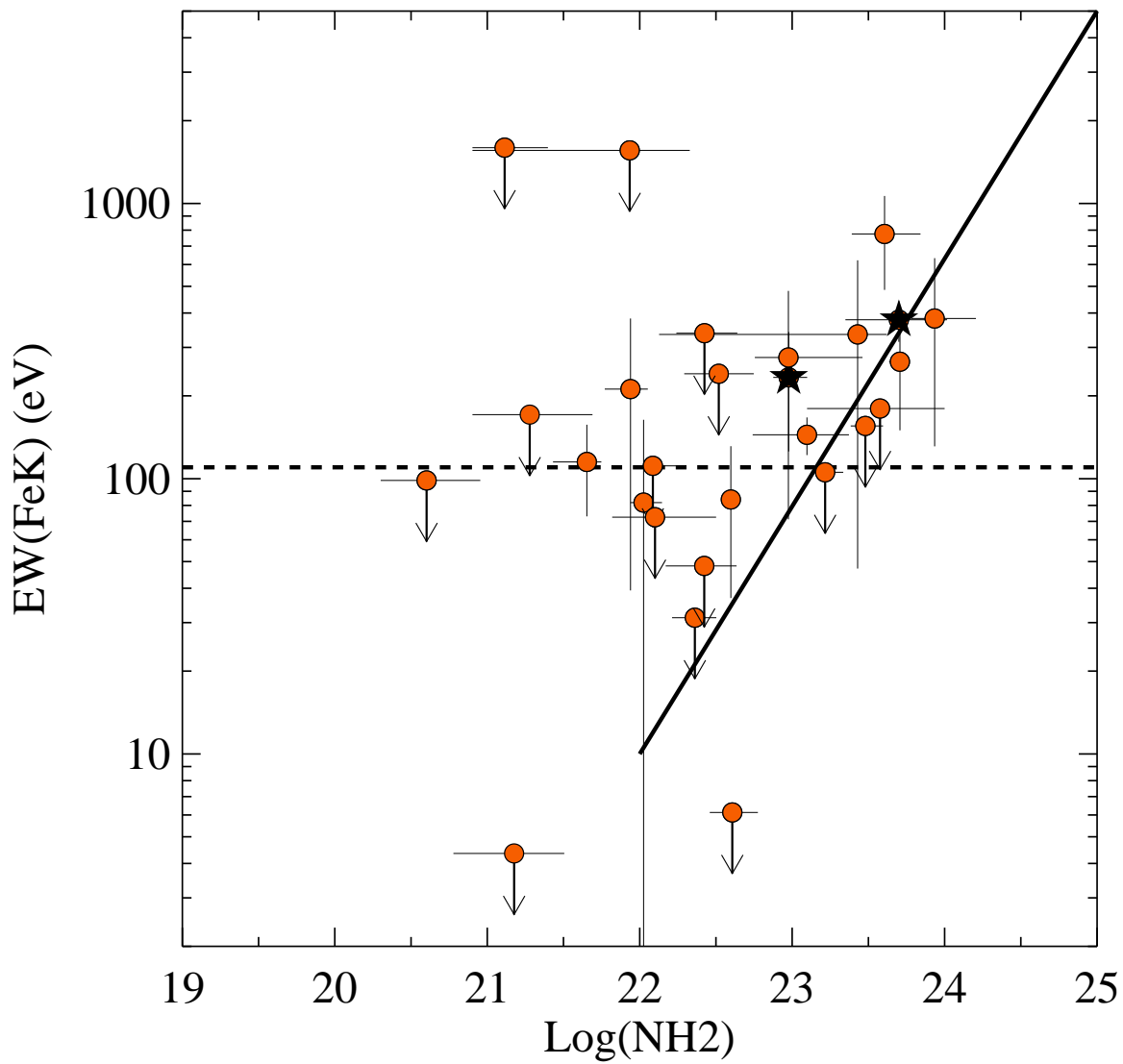}
\caption{EW of the FeK$\sf{\alpha}$ emission line versus the NH2
column density. Black stars are confirmed {\it Compton-thick} (see Sect. \ref{sec:published}).
The horizontal dotted line is the mean level reported by \citet{guainazzi_x-ray_2005}
for a sample of type 2 Seyferts. The diagonal continuous line
shows the predicted line reported by \citet{ghisellini_contribution_1994} for
attenuation by photo-absorption and {\it Compton}
scattering.}
\label{cha3:fig:EWNH2}
\end{center}
\end{figure}

In order to further explore the use of $\sf{EW(FeK\alpha)}$ as
a tracer for obscuration, we analyze whether its value depends on the
absorbing column density NH2 (Fig. \ref{cha3:fig:EWNH2}).  Upper
limits for column densities were excluded from this analysis. 
The trend that we find is consistent with the predictions by 
\citet{ghisellini_contribution_1994}\footnote{Below 
$\sf{10^{23}cm^{-2}}$, $\sf{EW(FeK\alpha)}$ seem to
remain constant with values around 100 eV and for NH2$\sf{>10^{23}cm^{-2}}$,
$\sf{EW(FeK\alpha)}$ appear to increase up to $\sf{\sim 500}$~eV at
$\sf{10^{24}cm^{-2}}$.}.

To validate the calculated values, we have also checked that they
do not depend on the choice of the X-ray spectral continuum. 
We have used two physical models to fit the underlying continuum 
in the {\it Compton-thick} scenario:
(1) A single reflection model to the hard X-rays (${\sf >2~keV}$) 
\citep[{\sc pexrav} in Xspec,][]{Magdziarz_1995} and 
(2) the {\it baseline} model for {\it Compton-thick} sources reported 
by \citet{guainazzi_x-ray_2005}.
Because of the LINER complex spectrum below 2 keV (fitted by 
thermal and/or power-law components), the first model could
add complementary information using only the 2-10 keV energy range.

We have used a pure reflection model by fixing the 
reflection scaling factor to ${\sf -1}$ and solar abundances.
The power-law spectral index is linked to the intrinsic 
continuum power-law in the second model.  
The fit is performed for nuclei with more than 200 counts 
in the band where the fit is made 
(i.e. 2-10 keV and 0.5-10 keV, respectively).
The resulting $\sf{EW(FeK\alpha)}$ are presented in Table \ref{cha3:tab2} (Col. 7 and 8).

In general, the EWs obtained 
using the best-fit reported by GM+09 are confirmed with these two models. 
Using the single reflection model above 2 keV and taking upper limits, NGC\,410 changes to  
$\sf{EW(FeK\alpha)>500~eV}$ and NGC\,4374 changes to $\sf{EW(FeK\alpha)<500~eV}$.
Using the baseline model for {\it Compton-thick}  sources, NGC\,0833 changes to 
$\sf{EW(FeK\alpha)}<$ $\sf{500~eV}$ and UGC\,05101 and NGC\,6240  show an upper 
value slightly higher than 500 eV (510 and 530 eV, respectively).

\section{Discussion}

\subsection{Previously found Compton-thick LINERs in our sample}\label{sec:published}

Based on {\it BeppoSAX} data, sensitive above 10 keV up to 300
keV, two of the LINERs in the sample, NGC\,3690B, and NGC\,6240, show intrinsic 
continuum above 10 keV, typical of {\it Compton-thick} sources
\citep{ptak_chandra_2003,ceca_enshrouded_2002,
risaliti_distribution_1999,vignali_probing_1999}. UGC\,05101 and
NGC\,5005 have been also claimed to be {\it Compton-thick} by using an
indirect method. Based on the detection of a high upper limit for
the $\sf{EW(FeK\alpha)}$ on \emph{ASCA} data
\citet{risaliti_distribution_1999} claimed a column density for NGC\,5005 larger
than 10$^{24}$ cm$^{-3}$. With more recent \xmm{} data, \citet{guainazzi_x-ray_2005} 
questioned its {\it Compton-thick} nature. A highly obscured nuclei 
(NH$\sim$10$^{24}$ cm$^{-3}$)  has been reported by \citet{imanishi_x-ray_2003} 
for UGC\,05101. Moreover, recently 
\citet{Teng_2008} have classified UGC\,08696 as a {\it Compton-thick} source,
using {\it Suzaku} data above 10 keV. However, we have decided not to use it as 
{\it Compton-thick} source because they have reported changes in the 
spectral shape that might be due to large changes in the absorbing column density. 
Hereinafter we consider UGC\,05101, NGC\,3690B, and NGC\,6240
objects as {\it Compton-thick} sources according to published literature results 
(hereinafter we refer to them as 'Confirmed {\it Compton-thick}').  For the
remaining nuclei in our sample, no signs of a {\it Compton-thick}
nature have been reported in the literature.

Concerning the indirect {\it Compton-thick} diagnostics used above, 
two of the confirmed {\it Compton-thick} show flat spectra (UGC\,05101 and
NGC\,6240) whereas NGC\,3690B  seems to have a somewhat
steeper spectrum. Thus, a classification of {\it Compton-thickness} based solely in
this criterion needs to be taken with some caution. Moreover, 
the three confirmed {\it Compton-thick} show  
$\sf{log(F_{X}(2-10~keV)/F([OIII]))<0}$, consistent with being 
{\it Compton-thick} object. Also, we have been able to determine the $\sf{EW(FeK\alpha)}$ of
the three confirmed {\it Compton-thick} in our
sample, finding $\sf{EW(FeK\alpha)=280\pm180~keV}$, $\sf{EW(FeK\alpha)=230\pm110}$ $\sf{keV}$ and
$\sf{EW(FeK\alpha)=380}$ $\sf{\pm 60}$ $\sf{keV}$, for  UGC\,05101, NGC\,3690B
and NGC\,6240 respectively (shown as black stars in
Figs. \ref{cha3:fig:EWNH2} and \ref{cha3:fig:LHOIII_fek}). These
values are compatible with the values 410$\sf{\pm 250}$~eV, 420 $\sf{\pm 260}$~eV
and 300$\sf{\pm 100}$~eV reported in the literature for UGC\,05101
\citep{imanishi_x-ray_2003}, NGC\,3690B \citep{ballo_arp_2004} and
NGC\,6240 \citep{boller_xmm-newton_2003}, respectively.  Our values do not agree
with those expected  for {\it Compton-thick} sources
($\sf{EW(FeK\alpha)}$ $\sf{>500~eV}$). When the {\it baseline} model for {\it Compton-thick} sources is
used, the values for UGC\,05101 and NGC\,6240 raise to 510 and 530 eV, respectively 
(although the fit is statistically worse than for the best fit model). We
discuss the use of a low EW(FeK$\sf{\alpha}$) for 
{\it Compton-thickness} diagnostics in Section \ref{sec:lowEW}.

\subsection{Compton-thick LINERs}

Table \ref{cha3:tab2} shows the {\it Compton-thick} classification following the
three tracers. Cols. 3, 5 and 9 of Table \ref{cha3:tab2} show the {\it Compton-thick}
classification according to the three diagnostics and Col. 10 gives the final  {\it Compton-thick}
classification. `CT' indicates {\it Compton-thick} classified, 
`?' not available information and `CT?' possible {\it Compton-thick} source.

We have information on at least one of the tracers for the whole sample. 
We have defined a source to be {\it Compton-thick} when
at least one tracer indicates such a classification and 
none of the others contradict it. 
When two tracers agree we have taken this classification, even 
if the third diagnostic does not agree. We have classified as `CT?'
four objects with only two tracers available showing contradicting results 
(NGC\,0833, NGC\,3690B, NGC\,5005, MRK\,266NE). However, note that
NGC\,3690B is a confirmed {\it Compton-thick} (see Sect. \ref{sec:published}).

Adding the information coming from the three {\it Compton-thick} tracers, 
{\it Compton-thickness} very often appears among LINERs, 
representing 49\% (40/82) of our sample (54\% including `CT?').  
Furthermore, at least one of the tracers indicates consistency with a {\it Compton-thick} nature
in 62\% of our sample. 

\citet{cappi_x-ray_2006} reported a sub-sample of 27 optically-selected and distance-limited Seyfert
galaxies (${\sf F(2-10 keV)>10^{13} erg~s^{-1}~cm^{-2}}$), 5 of them being {\it Compton-thick} (18\%). However, four
of their objects are in our LINER sample (NGC\,2685, NGC\,3185,
NGC\,4579 and NGC\,4698). Excluding these four objects, 4 out of the
remaining 23 objects are classified as {\it Compton-thick} candidates (17\%).
An extension of this sample was studied by \citet{panessa_x-ray_2006} 
(without distance completeness of the sample), founding arguments favouring the 
{\it Compton-thickness} nature in 11 out of their 47 Seyfert (23\%) galaxies; 
however, among them 10 objects are LINERs
(NGC\,2639, NGC\,2655, NGC\,2685, NGC\,3185, NGC\,3608, NGC\,3627,
NGC\,4579, NGC\,4698, NGC\,6482 and NGC\,7743). Excluding these
objects, 9 out of the 37 Seyfert galaxies in \citet{panessa_x-ray_2006} 
are {\it Compton-thick} sources (24\%).
For the 10 objects in common,  they classified
NGC\,3185 and NGC\,7743 as {\it Compton-thick} objects, in agreement
with our classification. However, we also classify as such NGC\,2639,
NGC\,2685, NGC\,3608, NGC\,3627 and  NGC\,4698. Note that their classification is based on 
the $\sf{F_{X}(2-10~keV)/F([OIII])}$ versus $\sf{F_{IR}/F([OIII])}$ diagram. All these 5 objects
show a $\sf{log(F_{X}(2-10~keV)/F([OIII]))<0}$ except NGC\,2639, that they classified as a
possible {\it Compton-thick} object. Closer to our findings, \citet{guainazzi_x-ray_2005} found a 
46\% {\it Compton-thick} 
sources in a sample of 49 nearby Seyfert galaxies (40\% if we exclude sources 
in our LINER sample). Thus, {\it Compton-thick} objects seems to be more frequently found in LINERs
than in type 2 Seyferts.

\subsection{Low $\sf{EW(FeK\alpha)}$ Compton-thick sources}
\label{sec:lowEW}

We have explored the validity of the $\sf{EW(FeK\alpha)}$ as a {\it Compton-thick}
tracer studying the connection between the $\sf{EW(FeK\alpha)}$ and
the NH2 column density (see Fig. \ref{cha3:fig:EWNH2}). 
Below $\sf{10^{23}cm^{-2}}$, the $\sf{EW(FeK\alpha)}$ of Seyfert 1 nuclei 
\citep{nandra_xmm-newton_2007} and unobscured quasars 
\citep{jimenez-bailon_xmm-newton_2005,piconcelli_2005}
seem to remain constant around 100 eV. For NH2$\sf{\gtrsim 10^{23}cm^{-2}}$,
the $\sf{EW(FeK\alpha)}$  seems to increase up to $\sf{\sim 500}$~eV at
$\sf{10^{24}cm^{-2}}$ for a sample of 49 type 2 Seyferts \citep{guainazzi_x-ray_2005}. 
Our trend does not seem to be inconsistent with this behaviour although 
better constraints on the EW, perhaps with new observations, 
would be required in order to make any conclusion.
The similarities of this trend with that reported by  \citet{guainazzi_x-ray_2005} 
would manifest that LINERs might be sharing the same
origin than Seyfert galaxies for the iron line, which is originated in the inner wall of the
torus \citep{ghisellini_contribution_1994}.

\begin{figure}
\begin{center}
\includegraphics[width=0.9\columnwidth,height=0.9\columnwidth]{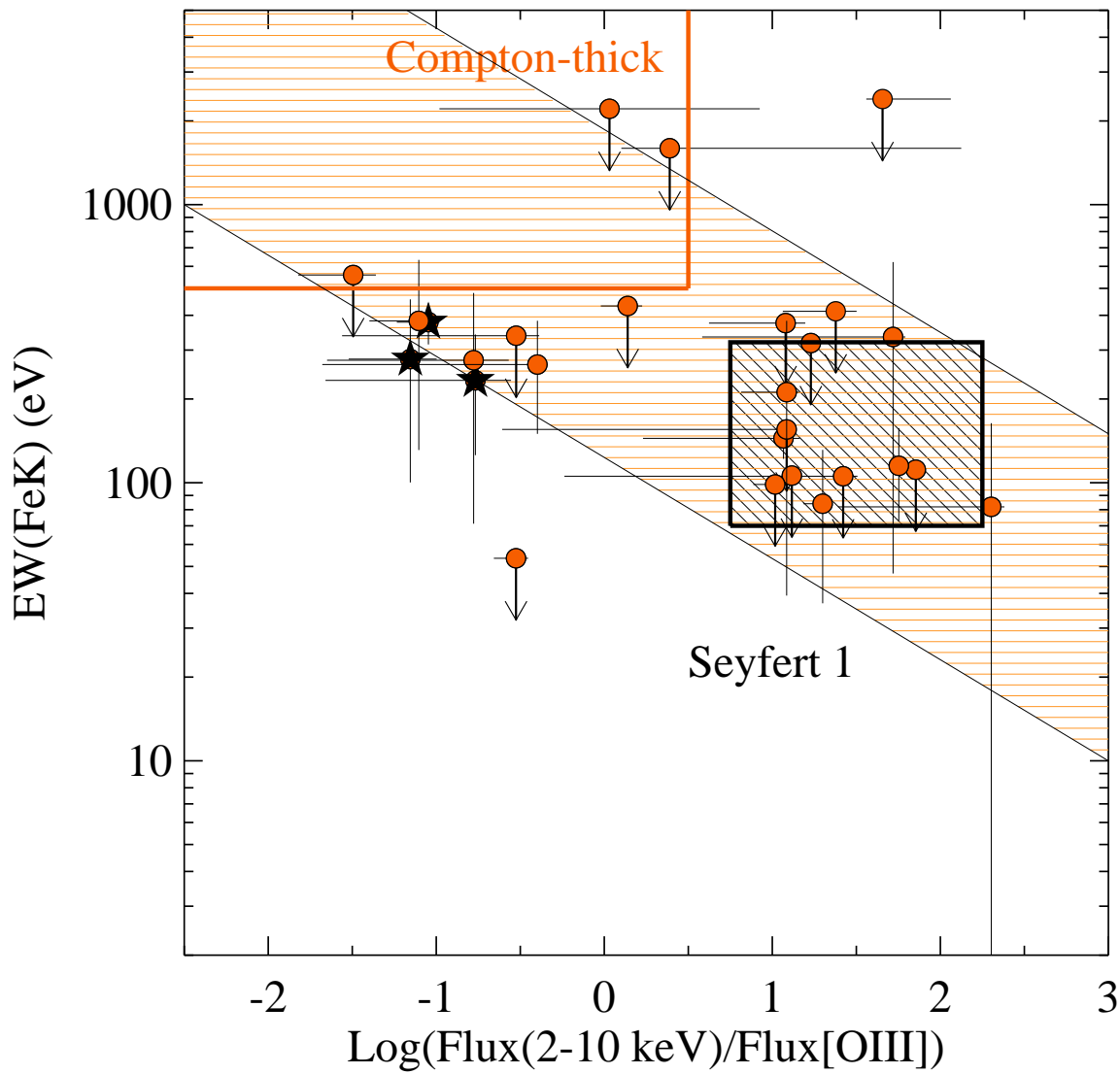}
\caption{EW of the FeK$\sf{\alpha}$ line versus the ratio between the
observed 2-10~keV flux and the dereddened [OIII] emission line flux. Black stars are
{\it Compton-thick} sources. The region filled with red horizontal lines shows the trend between this
two quantities shown by \citet{bassani_diagnostic_2000}. The square filled with diagonal black lines shows 
type 1 Seyferts location and the red-continuous line shows 
{\it Compton-thick} sources location \citep{bassani_diagnostic_2000}.}\label{cha3:fig:LHOIII_fek}
\end{center}
\end{figure}

Moreover, we have found a mismatch between the $\sf{EW(FeK\alpha)}$ and 
$\sf{log(F_{X}/F[OIII])}$ diagnostics. This is clearly seen in Fig. \ref{cha3:fig:LHOIII_fek}.
The correlation between these two quantities was used 
as a {\it Compton-thick} diagnostic by
\citet{bassani_three-dimensional_1999}. They found that both,
$\sf{EW(FeK\alpha)}$ and column density, decrease when the ratio
$\sf{F_{X}(2-10~keV)/F([OIII])}$ increases (region filled with red horizontal lines 
in Fig.\ref{cha3:fig:LHOIII_fek}). They found
$\sf{F_{X}(2-10~keV)/F([OIII])>0}$ and $\sf{EW(FeK\alpha)}$$\sim$100~eV for
type 1 Seyfert galaxies (square filled with diagonal black square), whereas {\it Compton-thick} galaxies were located
at $\sf{EW(FeK\alpha)}$ $\sf{>500~eV}$ and $\sf{F_{X}(2-10~keV)}$ $\sf{/F([OIII])<0}$ (red continuous line).

At least half of our sample is located in the region occupied by
type 1 Seyferts. However, neither of the confirmed 
{\it Compton-thick} sources (see Sect. \ref{sec:published}) fall in the expected region.
Two of them marginally fall in this regime 
when the {\it baseline} model for {\it Compton-thick} AGN is used.
We notice that in order to include
{\it Compton-thick} objects, the $\sf{EW(FeK\alpha)}$ limit needs to be
re-defined to a lower value of
$\sf{EW(FeK\alpha)}$$\sf{\sim}$200~eV. 
NGC\,2681, UGC\,05101, NGC\,3690B, NGC 4374,
NGC\,4410A, MRK\,266NE, NGC\,5363, NGC\,6240, IRAS\,17208-0014 and NGC\,7130 
fall in this regime. However, such a low limit on
$\sf{EW(FeK\alpha)}$ does not allow to distinguish between type 1
Seyferts and {\it Compton-thick} sources with low $\sf{F_{X}(2-10~keV)}$ $\sf{/F([OIII])}$.

Therefore, it seems that a new population of {\it Compton-thick} sources with low 
$\sf{EW(FeK\alpha)}$ emerges. The same 
result has been claimed by \citet{Brightman_2008}.
They explain their finding by postulating that the X-ray emission 
is dominated by scattering of the otherwise invisible obscured AGN emission, 
rather than by reflection from the inner wall of the torus.

{\it Compton-thick} LINERs are mostly fitted with a combination of a thermal plus
a power-law model. The later can be interpreted as the primary continuum or as the 
scattering component. Thus, the hard continuum can be completely lost under the scattering
continuum. However, the scattering component contributes always less than 10\% of the fluxes
above 2 keV for {\it Compton-thick} candidates. In LINERs, 
the thermal emission could be also responsible for the decreasing of the
$\sf{EW(FeK\alpha)}$ because a strong contribution of this component is found in a large 
number of them (see GM+09).

\subsection{Origin of the obscuring material}

In Seyfert 2 galaxies, the NH2 column density is related to the obscuration of the
primary continuum \citep{bianchi_x-ray_2004,guainazzi_x-ray_2005,
panessa_x-ray_2006,cappi_x-ray_2006}. This absorption in LINERs covers a range 
compatible with them. It does not appear to
be the case for NH1 column density:
consistent with the Galactic value in Seyferts (Bianchi et al., 2009),
NH1 column density in LINERs is an order of magnitude above the Galactic
value. The fact that our column density is larger than that
expected for Seyfert galaxies has opened the question of its
origin. 
Two possible origins arise: (1) material very close to the nucleus; 
(2) material in the host galaxies, with a column density 
larger than the expected for the Galactic value.

We have made the spectral analysis of the diffuse emission around the
nucleus using \cha{} data to investigate the origin of the NH1 column density.
 From the sample of 55 objects with spectral fitting,
we have selected the 
19 objects for which the spectral
analysis on the diffuse emission is expected to be reliable 
(see GM+09 for a detailed explanation of the diffuse emission analysis). 

\begin{figure}
\begin{center}
\includegraphics[width=0.9\columnwidth,height=0.9\columnwidth]{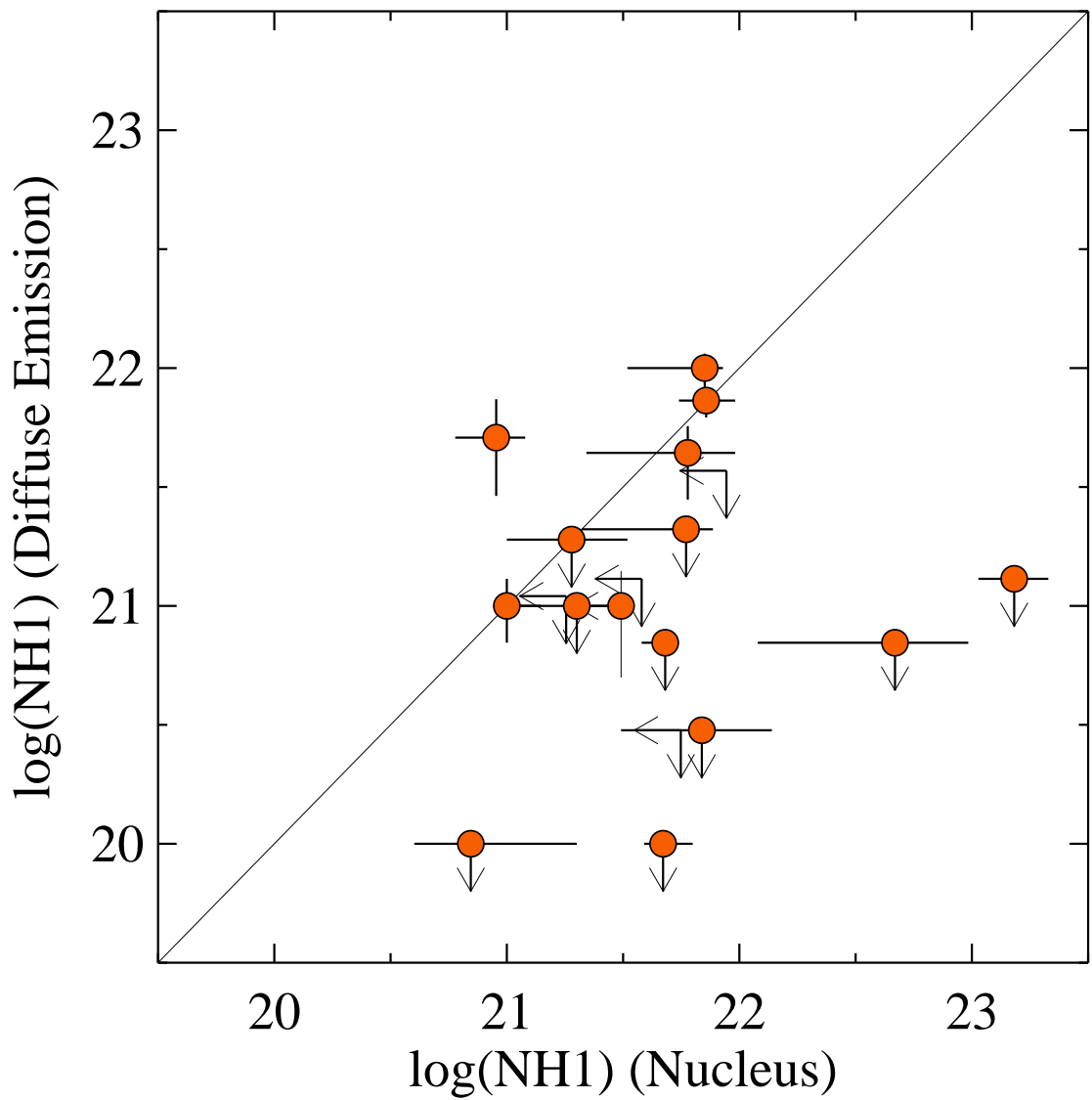}
\caption{NH1 column density of the diffuse emission (NH1(Diffuse Emission)) versus the 
NH1 column density of the nuclear emission (NH1(Nucleus)). Arrows are upper 
limits. The unity slope is shown as continuous line.}\label{cha3:fig:chadiffnhs}
\end{center}
\end{figure}

The comparison between the nuclear NH1 column densities (NH1(nucleus)) and
the diffuse emission NH1 column densities (NH1(diffuse emission)) is shown in
Fig. \ref{cha3:fig:chadiffnhs}. For six galaxies (NGC\,4125,
NGC\,4321, NGC\,4374, NGC\,4552, NGC\,4696 and IC\,1459) only upper
limits for NH1(diffuse emission) have been obtained.
In six galaxies (NGC\,0315, 3C\,218, NGC\,4111,
NGC\,4261, NGC 4579 and CGCG\,162-010) NH1(diffuse emission) is compatible with
the Galactic value, whereas NH1(nucleus) is larger. Thus, it can be safely
concluded that in these cases the material responsible for NH1 column density is unrelated 
to the host galaxy, and must be located in the nuclear region.
For NGC\,4278, NH1(diffuse emission) is larger than NH1(nucleus).
For the remaining 6 galaxies, nuclear and diffuse emission NH1 column densities lie, within the
errors, in the unity slope line, what can be interpreted as both
having the same origin, i.e. most probably related to the material 
in the host galaxy. 
In order to establish whether this is due to a dustier host 
galaxy environment in LINERs than in Seyferts, we discuss in the 
next subsection the optical Balmer decrement distribution in our 
LINER sample.

\subsection{X-ray versus optical obscuration}

\begin{figure*}
\begin{center}
\includegraphics[width=0.45\columnwidth]{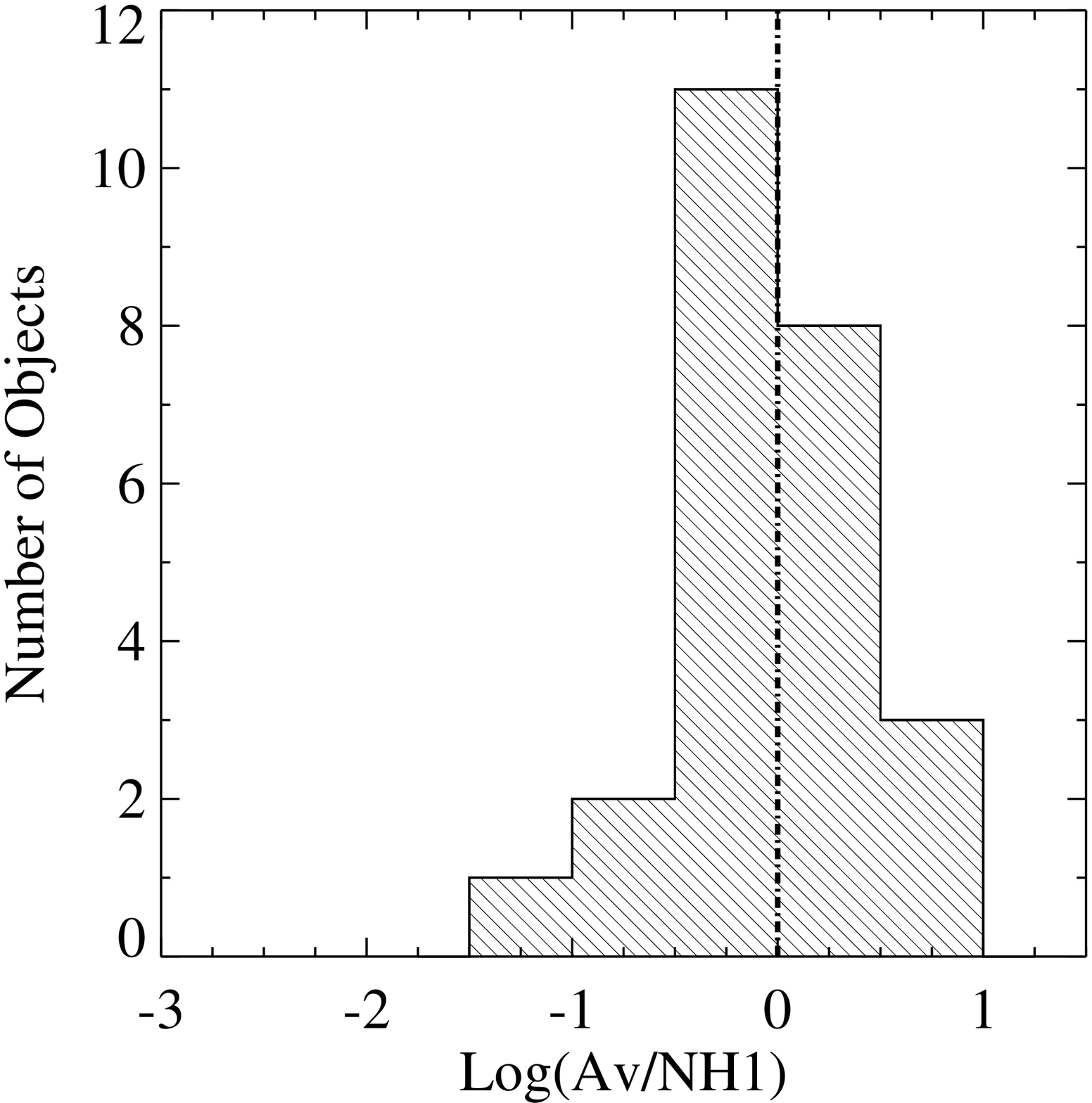}
\includegraphics[width=0.45\columnwidth]{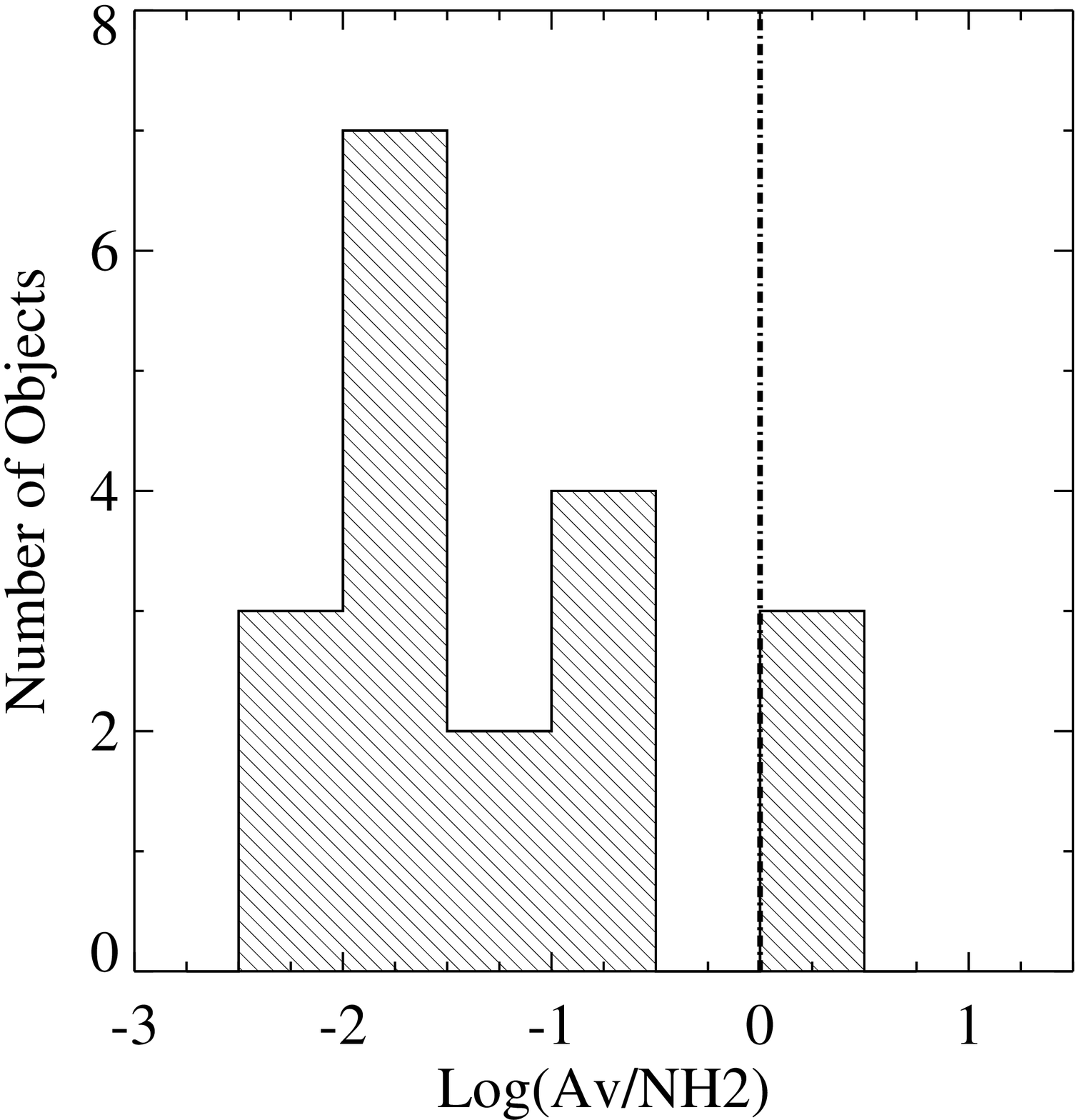}
\caption{Histograms of the ratio between the NH1 column density and optical extinction Av (\emph{Left})
and NH2 and Av (\emph{Right}). The dot-dashed line is the expected 
locus when the optical extinction Av can explain the X-ray column density (Av/NH=1). Optical extinction 
Av is converted into $\sf{cm^{-2}}$ units by assuming a 
Galactic ratio $\sf{Av/NH=5\times 10^{22}~cm^{-2}}$.}\label{cha3:fig:histoAVNH}
\end{center}
\end{figure*}

In this section we investigate the relation between the X-ray obscuring material 
and the optical extinction, Av. 
To do so, we have computed the optical extinction by using the
$\sf{H\alpha/H\beta}$ ratios. Objects showing the lower limit 
($\sf{H\alpha/H\beta=3.10}$) have been removed from this
analysis. Assuming a Galactic ratio $\sf{Av/N_{H}=5\times
10^{-22}cm^{-2}}$, we can express Av in units of $\sf{cm^{-2}}$. Fig.
\ref{cha3:fig:histoAVNH} shows the histograms of the ratio between
NH1 and Av (\emph{Left}) and 
NH2 and Av (\emph{Right}). 
While NH1 column density seems to be
related to the optical extinction, NH2 appears to be much larger.
This result was already drawn
by \citet{maiolino_obscured_2001} for a sample of type 2 Seyfert
galaxies. In their sample of unobscured type 2 Seyferts,
\citet{panessa_unabsorbed_2002} found that the obscuration
measured by the column densities at X-ray frequencies is consistent 
with that at the optical wavelengths (Av). This appears to be in good
agreement with our results, since the column density measured by
\citet{panessa_unabsorbed_2002} is more related to NH1 than to NH2
since they use only one absorption.

Interestingly, the overwhelming majority of LINERs in our sample 
(59 out of 67) show compact nuclear sources in {\it HST}\footnote{{\it HST} 
images, mainly with the filter F814W, can be seen
in the Appendix C (panel G) and GM+09.} (see Col. 9 of Table \ref{cha3:tab1}) 
sharp-divided images (see GM+06 for methodology explanation).
This is consistent with optical \emph{HST} data recently published by
\citet{gonzalez_delgado_hst/wfpc2_2008} for a sample of LINERs and
Transition Objects; we have found that only 5 out of their 34 LINERs
appear not to be compact. This rules out association of the optical 
dust line to the material responsible for NH1 column density.

Objects with a dusty environment (coded as 'D' in Table \ref{cha3:tab1}) 
are equally distributed between {\it Compton-thin} 
and {\it Compton-thick} sources. However, this result must be taken with some reserves
due to the low statistics. \citet{guainazzi_x-ray_2001}, in contrast to our result, found 
that {\it Compton-thin} Seyferts prefer dustier environment whereas 
{\it Compton-thick} Seyferts distribute both in dustier and
dust-free regions.
Based in the present data a relationship between
the optical morphology and {\it Compton-thickness} is not evident, being
LINERs not specially dustier objects.

\subsection{Obscuration and environment}

To take into account the eventual influence of the environmental
status in the properties of our sample galaxies, we have searched in
NED for possible companions at projected
distances smaller than 250 kpc. Galaxies have been classified into 4
groups according to their environment: 1) isolated, when no projected
companions with comparable sizes and redshift difference smaller than
1000 km/s are found, 2) pairs, when only another galaxy of comparable
size and close in redshift is identified (wide and close pairs
together with merging systems are grouped into this category), 3)
groups, for galaxies in small group environments (either compact or
loose), 4) clusters, for galaxies known to reside in cluster
environments (in some cases, they correspond to the cluster center,
see GM+09). 

\begin{figure}
\begin{center}
\includegraphics[width=0.9\columnwidth]{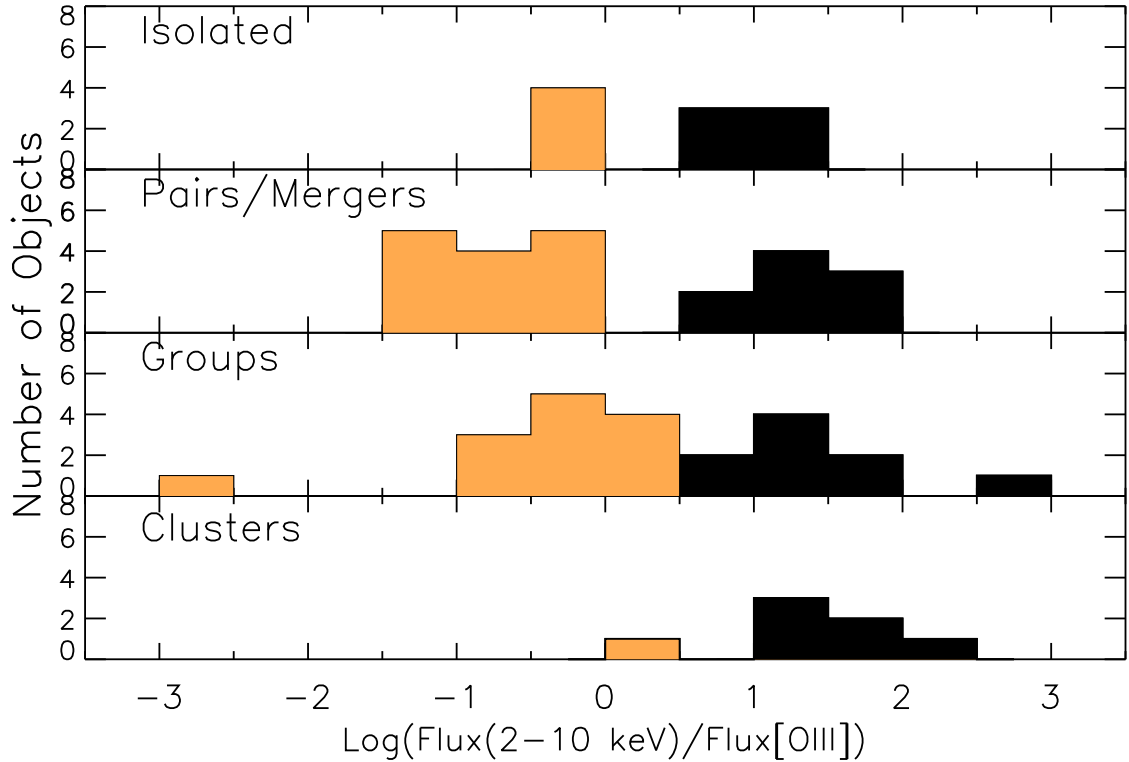}
\caption{Histogram for the L(2-10 keV)/L([OIII]) for the different environments: 
isolated (a), pairs and mergers (b), groups (c) and clusters (d), from top to 
bottom. Red filled histograms are {\it Compton-thick} sources  and black-filled 
histograms are {\it Compton-thin} sources (according to the  
L(2-10 keV)/L([OIII]) ratio).}\label{cha3:fig:hist_CT-envi}
\end{center}
\end{figure}

The environmental status does not seem to be connected
with any of the analyzed properties, except the {\it Compton-thickness}.  
The resulting average and sigma
values of ${\sf L(2-10 keV)}$ $\sf{/L([OIII])}$ for the different environments are provided
in Table \ref{cha3:tab5}, and the corresponding histograms
shown in Fig. \ref{cha3:fig:hist_CT-envi}. 
We note that {\it Compton-thick} objects seem to be frequent in cluster environments (14\%), but are
relatively frequent in groups (67\%), pairs and merging systems (65\%). 
The probability that cluster and group distributions are the same is 19\%.

This result could be understood by considering that a huge amount of
obscuring material is being transported to the nuclear regions as a
result of the merging process \citep[for
instance][]{mihos_h_1994,barnes_h_1996}.  In fact, for ULIRGs, galaxy
mergers are invoked to produce massive in-falls of gaseous material
towards their centers \citep[e.g.][]{rupke+08} and column densities $\ge$
10$^{24}$cm$^{-2}$ have been deduced from CO measurements
\citep[e.g.][]{downes+98,evans+02}.

Nevertheless, a number of galaxies in the group of
isolated objects appear 
also as {\it Compton-thick} candidates (NGC\,2639, NGC\,2685, NGC\,4457, 
NGC\,4698, NGC\,7331 and NGC\,7743). The origin of the obscuring material
in these systems has to be related either to intrinsic properties or
to secular processes in their host galaxies.

The statistical results are not conclusive and 
a bigger sample would be needed with a better coverage of the different environments.
However, the results are suggestive of a 
connection between the environment and the {\it Compton-thickness}.
\citet{Krongold_2003} found that type 2 Seyferts and type 2 LINERs show
the same rate of companions in the local universe, while type 1 Seyferts show a
lower rate of companions. However, recently \citet{Montero-Dorta_2008} have found that high 
redshift LINERs in DEEP2 tend to favour higher density environments while 
Seyfert do not show environmental dependencies. 

If the fraction of {\it Compton-thick} LINERs is dependent 
on the environment, a large population of {\it Compton-thick}
sources might be existing in the high redshift Universe.
This is the case of IRAS\,00182-7112, an optically classified LINER nucleus at redshift 0.3. 
The FeK$\sf{\alpha}$ emission line is not present, consistent with our nearby LINER sample,   
but a high EW of the He-like Fe K line and a flat spectrum indicate that the 
source is reflection-dominated \citep{Nandra_2007}.
Unfortunately, this population is probably missed with the current 
instrumentation, due to their low luminosity.

\subsection{Obscuration and Eddington ratios}\label{sec:eddington}

Several authors have pointed out that the difference between LINERs 
and Seyferts comes from their Eddington ratios being smaller in LINERs
\citep{dudik_chandra_2005,ho_new_2008}. 
Here we include the {\it Compton-thick} nature of the sources
into the analysis. We will demonstrate the importance of their 
effect in the estimation of the Eddington rate.

Assuming that all the candidates 
are actually {\it Compton-thick}, their intrinsic column
densities should be higher than $\sf{10^{24}cm^{-2}}$. Hence, their
X-ray luminosities are underestimated since the applied column density
correction is lower than the real. To estimate the correction factor  for 
{\it Compton-thick} sources
we divided our objects in two groups with
$\sf{F_{X}}$ $\sf{/F([OIII])>0.5}$ ({\it Compton-thin}) and
$\sf{F_{X}}$ $\sf{/F([OIII])<0.5}$ ({\it Compton-thick}),
respectively.  The median values are 
$\sf{<F_{X}}$ $\sf{/F([OIII])_{thin}>}$ = 16.5 and
$\sf{<F_{X}}$ $\sf{/F([OIII])_{thick}>}$ = 0.4. We use
the ratio between these two quantities as the correction
factor between {\it Compton-thick} and {\it Compton-thin} sources:
$\sf{<F_{X}}$ $\sf{(2-10}$ $\sf{ keV)/F([OIII])_{thin}>}$
$\sf{/<F_{X}}$ $\sf{(2-10}$ $\sf{ keV)/F([OIII])_{thick}>}$ =41.4
\footnote{Note that the same method was applied by \citet{panessa_x-ray_2006} to correct 
{\it Compton-thick} Seyferts, using as the reference value that for Type 1 Seyferts.}.
Applying this correction factor to our {\it Compton-thick} sources we
can obtain the corresponding corrected X-ray luminosities 
(reported in Table \ref{cha3:tab1}, Col. 6). 
The median value and standard deviation of the hard X-ray luminosity and of the ratio between
the {\it Compton-thick} corrected (2-10 keV) and [OIII] emission line luminosities is now 
log(L([OIII]))=40.5$\rm{\pm}$1.4 and $\sf{F_{X}(2-10~keV)/F([OIII])}$ = 1.1$\rm{\pm}$0.8, respectively.

\begin{figure*}
\begin{center}
\includegraphics[width=0.45\textwidth,height=0.9\textwidth,angle=90]{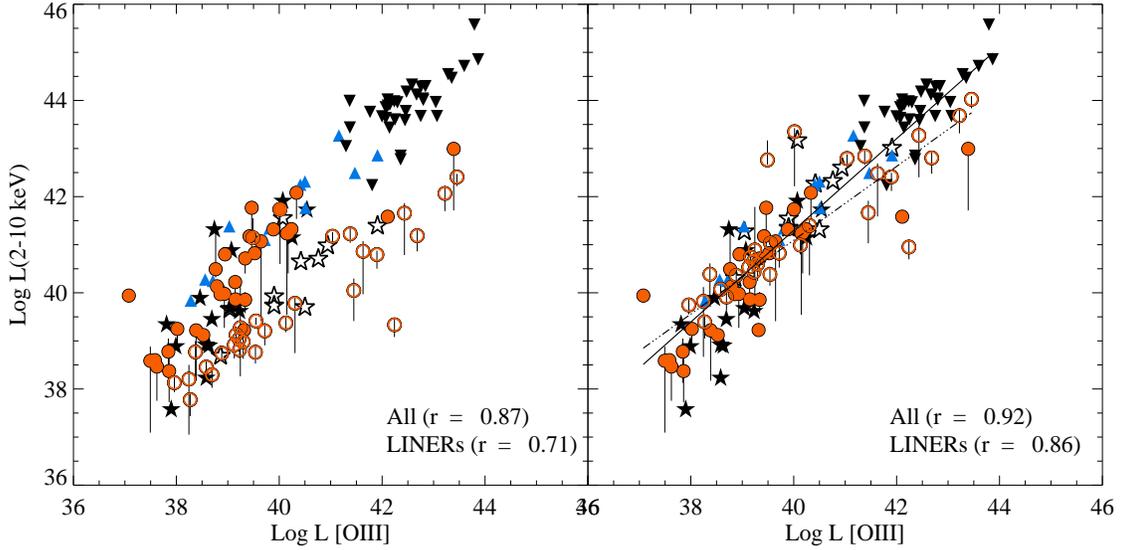}
\caption{L(2-10~keV) versus L([OIII])  before
(\emph{left}) and after (\emph{right}) the {\it Compton-thickness}
correction. Red-filled circles are {\it Compton-thin} LINERs, open circles correspond to {\it Compton-thick}
candidates within our LINER sample, black-filled stars are {\it Compton-thin} Seyferts, 
black-open stars are {\it Compton-thick} Seyferts, blue triangles are type 1 Seyferts 
\citep[Seyferts taken from by ][]{panessa_x-ray_2006}, upside-down black triangles are PG QSOs reported by 
\citet{jimenez-bailon_xmm-newton_2005} and \citet{piconcelli_2005}. The linear fit for LINER nuclei is 
shown as a dot-dashed line while the fit for all the AGN families
is shown as a continuous line. The correlation coefficient (r) is given in the 
bottom-right text  of each panel.}\label{cha3:fig:LHLOIII}
\end{center}
\end{figure*}

Fig. \ref{cha3:fig:LHLOIII} shows the hard X-ray (2-10~keV)
versus the [OIII] emission line luminosities before (\emph{Left}) and after
(\emph{Right}) the {\it Compton-thickness} correction. 
For LINERs, the former quantities are related by (circles in Fig. \ref{cha3:fig:LHLOIII}):

\begin{equation}
log~L_{X}(2-10 keV)=(10\pm2)+(0.8\pm0.1)~ log~L([OIII])
\end{equation}

Note that \emph{Compton-thin} and \emph{Compton-thick} LINERs populate the same location than 
\emph{Compton-thin} and {\it Compton-thick} Seyferts, respectively. Moreover, unobscured 
QSOs nicely extend to the brightest regime of the correlation. Note that there is a lack of high [OIII] luminosities
seen in {\it Compton-thin} LINERs compared with {\it Compton-thick} LINERs.
This is due to a selection effect since we are selecting LLAGN.

After {\it Compton-thick} corrections
all the families show a linear correlation: 

\begin{equation}
log~L_{X}(2-10 keV)=(3.2\pm1.4)+(0.95\pm0.03)~ log~L([OIII])
\end{equation}

This correlation agrees with that recently found by \citet{Lamastra_2009} for a sample of {\it Compton-thin}
type 2 Seyferts. The good correlation of all the families of AGN, including LINERs (correlation coefficient of r=0.92), 
is an argument favoring the same emission mechanism for all the objects.

\citet{dudik_chandra_2005} have found an Eddington ratio			
$\sf{L_{bol}/L_{Edd}=7\times 10^{-6}}$ for a sample of LINERs. These		
low accretion rates would indicate an inefficient				
accretion flow \citep[$\sf{<10^{-3}}$ defined in][]{terashima_x-ray_2002}.	
The Eddington ratio is defined as $\sf{L_{bol}/L_{Edd}}$.  Since the		
bolometric luminosity, $\sf{L_{bol}}$, estimated as in  			
\citet{dudik_chandra_2005}, is directly related to the hard X-ray luminosity	
as $\sf{L_{bol}}$ = 34$\sf{\times}$L(2-10~keV), 				
{\it Compton-thickness} correction may have an impact in the derived values of the
Eddington ratios and hence in their interpretation.				
										
We have revisited this question in LINERs by calculating their Eddington	
ratios with the {\it Compton-thick} corrected luminosities. Considering 	
${\sf L_{Edd}}$ = 1.26 x 10${\sf ^{38}}$					
M${\sf _{BH}}$/M${\sf_{\odot}}$ erg s${\sf ^{-1}}$ \citep{peterson_1997},	
we have estimated the black-hole masses by using the correlation		
between stellar velocity dispersion and black hole mass reported by		
\citet{tremaine_slope_2002}:							
										
\begin{equation}								
log (M_{BH}(M\odot)) = 8.13 +  4.02 ~ log(\sigma (km~s^{-1})/200)		
\end{equation}  								
										
\begin{figure}
\begin{center}
\includegraphics[width=0.9\columnwidth,height=0.9\columnwidth]{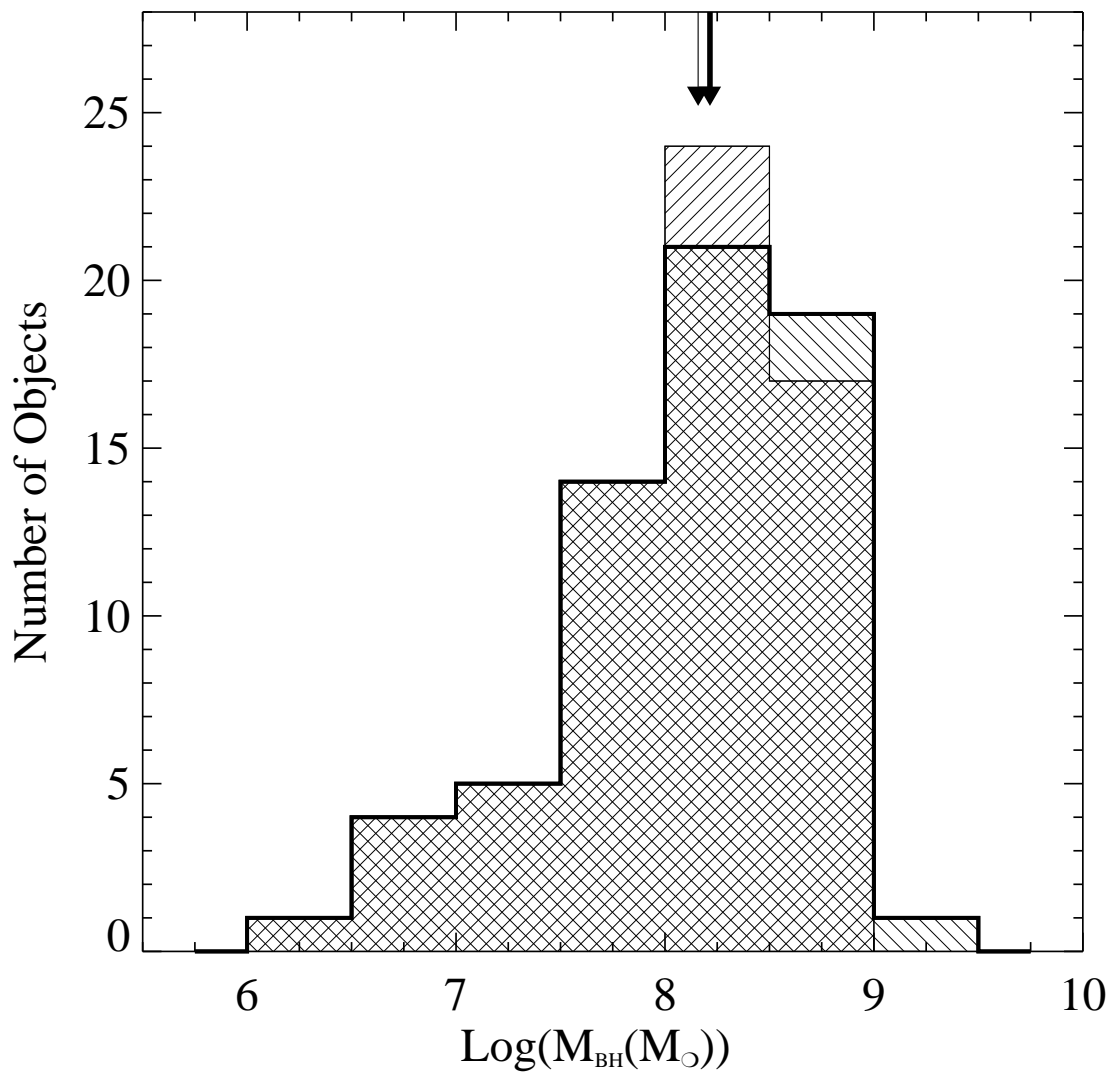}
\caption{Histogram for black hole masses. The thick-filled region (top-left to bottom-right lines)
shows the distribution obtained using the relation by  \citet{tremaine_slope_2002} and the thin filled region (bottom-left to top-right lines) show 
the distribution using that by \citet{Ferrarese_2001}. Black thick and thin arrows show the locii of the median value using 
\citet{tremaine_slope_2002} and \citet{Ferrarese_2001}, respectively. }\label{cha3:fig:hist_BH-tipo}
\end{center}
\end{figure}

The velocity dispersions have been taken from					
Hyperleda\footnote{http://leda.univ-lyon1.fr/}, where they are available	
for 63 out of the 82 objects (Table \ref{cha3:tab1}, Cols. 10 and 11).  	
The distribution of black hole masses is plotted in				
Fig. \ref{cha3:fig:hist_BH-tipo} (Dashed from up-left to down-right region).	
They range from $\sf{log(M_{BH})=6}$						
to  $\sf{log(M_{BH})=9.5}$ with a median value of	
$\sf{log(M_{BH})}=$ 8.22$\rm{\pm}$ (and standard deviation of 0.65).					
Our values agree with those for the nine objects in common with 		
\citet{Walsh_2008}, with STIS multi-spec data.  				
Dudik et al. (2005) derived the black hole masses following		
\citet{Ferrarese_2001} instead of  the above equation. For comparison purposes  
we have also estimated the black-hole masses using this equation (Fig.  	
\ref{cha3:fig:hist_BH-tipo}, dashed from down-left to up-right region). 	
Using \citet{Ferrarese_2001} we derive a median value of		
$\sf{log(M_{BH})}=$ 8.16$\rm{\pm}$ (and standard deviation of 0.61), compatible with the values		
reported by using the approximation by \citet{tremaine_slope_2002}.

\begin{figure}
\begin{center}
\includegraphics[width=0.45\textwidth,height=0.9\textwidth,angle=90]{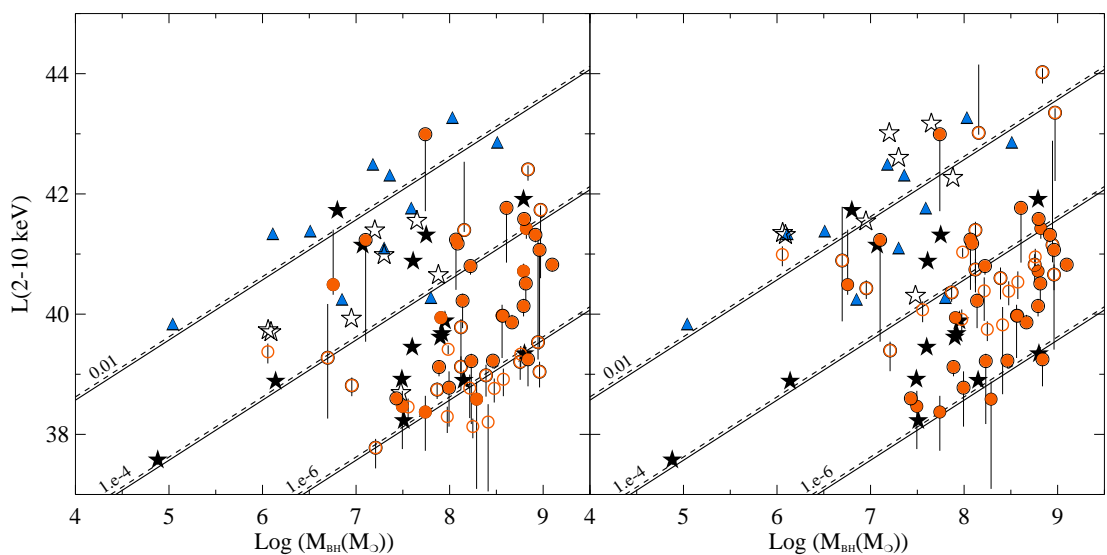}
\caption{Hard X-rays (2-10~keV) luminosities versus the black hole masses, before      
(\emph{Left}) and after (\emph{Right}) {\it Compton-thickness} correction.	       
Symbols as in Fig. \ref{cha3:fig:LHLOIII}. The solid and dashed lines show the L(2-10 keV) luminosity as a function of      
the black hole mass for Eddington ratios of 0.01, $\sf{10^{-4}}$  and $\sf{10^{-6}}$
\citep[assuming $\sf{L_{bol}/L(2-10~keV)= 34}$ and $\sf{L_{bol}/L(2-10~keV)= 30}$,][respectively]{risaliti_panchromatic_2004,panessa_x-ray_2006}.}		       
\label{cha3:fig:BHM_LX}
\end{center}
\end{figure}

Fig. \ref{cha3:fig:BHM_LX} shows L(2-10~keV) versus the black			
hole mass, before (\emph{Left}) and after (\emph{Right})			
{\it Compton-thickness} correction. The data reported for			
type 1 and 2 Seyferts by \citet{panessa_x-ray_2006} have been included.
LINER black hole masses tend to be larger than those for  		
type 2 Seyferts reported by \citet{panessa_x-ray_2006}  			
(average log(${\sf M_{BH}}$)= 7.0). \citet{panessa_x-ray_2006} took their  
black hole masses from the literature, from kinematics methods to		
reverberation mapping or from the mass-velocity dispersion correlations.	
However, we do not expect any noticeable difference coming from this		
assumptions because our LINER black hole mass estimations are consistent	
with other methods (See Fig. \ref{cha3:fig:hist_BH-tipo}).	
We have verified that the difference between black hole masses of Seyferts and LINERs  
can be attributed to the different distribution in morphological types, LINERs  
being hosted by earlier morphological type galaxies.

Before the {\it Compton-thickness} correction the median Eddington		
ratio was $\sf{L_{bol}/L_{Edd}=}$ $\sf{3.2\times 10^{-6}}$			
\citep[$\sf{L_{bol}/L_{Edd}=}$ $\sf{2.8\times 10^{-6}}$, assuming the value of		
Lbol=30$\rm{\times}$L(2-10 keV), taken from][]{cappi_x-ray_2006}, 	
consistent with the result by \citet{dudik_chandra_2005}. Nevertheless, after	
the {\it Compton-thickness} correction, this ratio increases up to		
$\sf{L_{bol}/L_{Edd}=1.9\times 10^{-5}}$ ($\sf{L_{bol}/L_{Edd}=}$  	
$\sf{1.8\times 10^{-5}}$, assuming Lbol=30$\rm{\times}$L(2-10 keV)).		
Type 2 Seyfert galaxies by \citet{cappi_x-ray_2006} 
\citep[also ][]{panessa_x-ray_2006} cover the same range of Eddington ratios,		
although they tend to be located close to						
$\sf{L_{bol}/L_{Edd}\simeq 1\times 10^{-3}}$. Type 1 Seyferts tend to be
located at higher Eddington ratios.
LINERs tend to be located closer to						
$\sf{L_{bol}/L_{Edd}\simeq 1\times 10^{-5}}$.					
The fact that Seyferts appear in later morphological types, typically		
Sab against S0 in LINERs, could explain the observed trend.			
										
Radiatively inefficient accretion is invoked to explain such low  	
Eddington ratios. \citet{merloni_2003} argued that the accretion		
mode changes to radiatively inneficient process below $\rm{10^{-3}}$. 	
Following it, all but seven sources are in the regime of radiatively  	
inneficient process. Therefore, for the overall picture of LINERs, 
their $\sf{L_{bol}/L_{Edd}}$ ratio, as well as for the type 2 Seyferts reported by \citet{panessa_x-ray_2006}, 
is within the inefficient regime. Therefore, a combination of obscuration 
and inefficient accretion is needed to explain their properties.
LINERs, therefore, share the range of accretion rates and 
obscuration of type 2 Seyferts, but occupying lower accretion rates and 
higher obscurations than their parent population of Seyferts.				

\section{Obscuration and SED}

Ho et al. (1999) (see Ho 2008 for a review) showed that the SED of LLAGN with 
$\sf{L_{bol}/L_{Edd}<1\times 10^{-3}}$ emphasizing that the big blue bump
seen in luminous AGN is absent in these sources, shifting the peak to 
the mid IR frequencies. Two possibilities exist: (1) Their nuclei are obscured 
and their UV photons are re-emit toward mid-IR 
frequencies or (2) they have a inherent different SED. 

\citet{ho_new_2008} ruled out the obscuration possibility mainly 
because of the lack of obscuration seen at X-ray in LINERs. However, in this study 
(also GM+09) we have demonstrated that they show X-ray obscuration 
(similar distribution of column densities than type 2 Seyferts) and a high 
percentage of them could be highly obscured AGN (higher percentage than type 
2 Seyferts in the {\it Compton-thick} regime). Therefore, the 
unobstructed view of the nucleus discussed by Ho (2008) is not so clear 
and obscuration of their nuclei becomes again relevant.

In the case of a intrinsically different SED, 
this could produce a lack of UV photons and, thus, a faint
production of the [OIII] emission. Therefore, this effect could affect 
those objects with large $\sf{F_{X}(2-10~keV)/F([OIII])}$. 
Thirty seven objects showed a high ratio ($\sf{log(F_{X}(2-10~keV)/F([OIII]))}$ $\rm{> 0.5}$). 
However, it is worth noticing that the same effect can be produced by a
wrong estimation of the extinction (discussed in Section 2.2). 
Eleven out of these 37 objects were not corrected from extinction, assuming a 
minimum of $\rm{H\alpha/H\beta}=$3.1. Excluding them, 26 objects showed 
l$\sf{log(F_{X}(2-10~keV)/F([OIII]))}$ $\rm{> 0.5}$. However, a hydrogen column density higher 
than $\rm{NH>10^{21}~cm^{-2}}$ is present in all but 5 cases (NGC\,3945, 
NGC\,4636, NGC\,4696, IRAS\,14348-1447 and MRK\,848). In these five 
cases an intrinsic different SED might be the only explanation for their 
large $\sf{F_{X}(2-10~keV)/F([OIII])}$. However, this is not the generality of the sample, 
in which the obscuration is a very important ingredient. 

In fact, \citet{Maoz_2007} reported the X-ray-to-UV ratio ($\rm{\alpha_{ox}}$) 
for a sample of 13 LINERs. He found a ratio between -1.4  $\rm{<\alpha_{ox}<}$ -0.8.
Seven out of the 13 objects are included in our sample. All of them show ratios 
$\sf{F_{X}(2-10~keV)/F([OIII])}$ $\rm{> 0.5}$, as expected if these objects could show a deficit
of UV photons. However, their NH column densities are
higher than $\rm{NH>10^{21}~cm^{-2}}$, that  
could also affect the UV band. Intriguingly, in spite of the lack 
of statistic (only seven objects), we have found a hint of correlation 
between $\rm{\alpha_{ox}}$ and the NH column density, obtaining a 
coefficient of a correlation of r=0.56. NGC\,5494 is clearly out of the 
correlation (excluding it, we obtain r=0.80). A bigger sample of LINERs 
with hydrogen column densities as well as X-ray and UV luminosities 
would be needed before any strong conclusions can be made.

For the whole energy distribution, \citet{Nemmen_2006} studied the LINER 
NGC\,1097 SED, successfully reproducing using a model of an optically thin geometrically thick
RIAF for the inner radii and an optically thick, geometrically thin disk for larger radii.
They needed to add a contribution of young and obscured 
starburst to account for the UV excess emission. However, although this is the case for NGC\,1097, 
we know that young circumnuclear starburst are not an important ingredient in most of LINERs 
(see GM+09). We need to study the LINER SED in a larger sample of objects to check if they can be 
explained with an ADAF model and the obscuration reported here.

\section{Summary and conclusions}

In this paper, we study the obscuration in LINER nuclei. We use 
X-ray spectral parameters, FeK${\sf \alpha}$ emission line and 
luminosities reported in GM+09 and in this paper. We also make use of their
optical properties traced by {\it HST} morphology, [OIII] emission line fluxes, optical extinction
and environmental information. 

Our key finding is that around 50\% of our LINER sample shows {\it Compton-thick} signatures, according to
the accessible diagnostics. This fraction is larger than that reported for Seyfert nuclei.
Moreover, this high percentage of LINERs showing high obscuration consistent 
with the observed decrease of obscured AGN fraction with luminosity \citep{DellaCeca_2008}.

This population of {\it Compton-thick} LINERs shows lower 
$\sf{EW(FeK\alpha)}$. \citet{Brightman_2008} suggest 
that the scattered emission dominating below $\sf{\sim 5~keV}$, what leads to a misinterpretation of the spectrum as an unobscured object
while a nuclear harder component is present. As pointed by \citet{Brightman_2008}, the 
high contribution of the scattering component might suggest a low covering fraction of the torus or a 
high density of the gas. The thermal emission in LINER nuclei, however, seems to be another 
important ingredient since it is needed in the majority of LINERs, 
accounting for a high fraction of the emission above $\sim 2~keV$.
X-ray spectroscopic measurements above 10 keV together with high 
sensitivity spectra around 6.4 keV are required for a better understanding of their nature.

We have also investigated the connection between hydrogen column densities and 
the optical extinction and environment. The soft column density (NH1) is related with 
the diffuse emission around the AGN at least in some cases, and it correlates with the optical extinction.
The hard column density (NH2) is much higher than the optical extinction, that might be   
associated to the inner parts of the AGN, buried at optical wavelengths.
No relation was found between optical dust lanes and
X-ray obscuration or 
{\it Compton-thick} LINERs. 
Finally, LINER nuclei show lower 
Eddington ratios than type 2 Seyferts, although they cover the same range 
of values. We want to remark that the {\it Compton-thickness} luminosity correction is very important 
for a proper estimation of their Eddington ratios.

\vspace{1cm}

Therefore, obscuration plays an important role in LINERs. Close to our findings, 
\citet{Dudik_2008} found that a high extinction 
even at mid-IR frequencies is needed to explain the deficit of NeV $\rm{\lambda \lambda 14~\mu}$m 
compared with NeV $\rm{\lambda \lambda 24~\mu}$m. In fact,
their seven objects in common with our sample (NGC\,1052, NGC\,3507, NGC\,4736, 
UGC05101, UGC\,08696, NGC\,6240 and NGC\,7130) with deficit of NeV $\rm{\lambda \lambda 14~\mu}$m 
show a high NH column density or have been classified as {\it Compton-thick} sources.
High obscuration of the central source might also explain the lack of the UV bump in 
LINERs.

Our results seem to be consistent with an scenario in which LINER nuclei are characterized by 
two phases of obscuration. 
The hard X-ray obscuring material will be similar to that obtained 
for type 2 Seyferts and will be much more obscured than that 
expected to produce the optical
extinction. A possibility that needs to be investigated is whether it is related to
the putative dusty torus invoked by the unified model for AGN nuclei.
\citet{nenkova_dust_2002} proposed that the obscuring region in AGN
is a toroidal distribution of dusty clouds.
\citet{ibar_constraining_2007} predicted an intrinsic {\it
Compton-thick} source fraction of 58\% for the clumpy torus scenario while the `classical' torus
produces a fraction of {\it Compton-thick} sources of 27\%. Our results
are closer to the expectations for a clumpy torus. 
Outside the hard X-ray absorbing material, LINERs might also show a second phase
of obscuration. This second phase will be responsible for the optical extinction, 
and will completely bury the intrinsic continuum at optical wavelengths.

\section*{Acknowledgements}
We acknowledge the referee for his/her helpful comments. We also thank D. Dultzin 
for fruitful discussions. 
We gratefully acknowledge F. Durret, J. Acosta, F. Carrera and 
E. Florido, members of OGM's PhD jury. We also thank to A. de 
Ugarte Postigo for the help to improve the text.  
This work was financed by DGICyT grants AYA
2003-00128, AYA 2006-01325, AYA 2007-62190 and the Junta de
Andaluc\'{\i}a TIC114. OGM acknowledges financial support from the
Ministerio de Educaci\'on y Ciencia through the Spanish grant FPI
BES-2004-5044 and research fellowship of STFC. This research has made use of data obtained from the
\emph{Chandra} Data Archive provided by the \emph{Chandra} X-ray Center (CXC) 
\emph{XMM-Newton} Data Archive provided by the \emph{XMM-Newton} Science Archive (XSA).
This paper is also partially based on NASA/ESA
{\it Hubble Space Telescope} observations. This
research made use of the NASA/IPAC extragalactic database (NED), which
is operated by the Jet Propulsion Laboratory under contract with the
National Aeronautics and Space Administration, and Hyperleda
database. 
 

\begin{thebibliography}{}

\bibitem[\protect\astroncite{Antonucci}{1993}]{antonucci_unied_1993}
Antonucci, R.: 1993,
\newblock {\em Annual Review of Astronomy and Astrophysics} {\bf 31}, 473

\bibitem[\protect\astroncite{Ballo et~al.}{2004}]{ballo_arp_2004}
Ballo, L., Braito, V., Ceca, R.~D., Maraschi, L., Tavecchio, F., and Dadina,
  M.: 2004,
\newblock {\em Astrophysical Journal} {\bf 600}, 634

\bibitem[\protect\astroncite{{Barnes} and {Hernquist}}{1996}]{barnes_h_1996}
{Barnes}, J.~E. and {Hernquist}, L.: 1996,
\newblock {\em Astrophysical Journal} {\bf 471}, 115

\bibitem[\protect\astroncite{Bassani
  et~al.}{1999}]{bassani_three-dimensional_1999}
Bassani, L., Dadina, M., Maiolino, R., Salvati, M., Risaliti, G., della Ceca,
  R., Matt, G., and Zamorani, G.: 1999,
\newblock {\em Astrophysical Journal Supplement Series} {\bf 121}, 473

\bibitem[\protect\astroncite{Bassani et~al.}{2000}]{bassani_diagnostic_2000}
Bassani, L., Dadina, M., Maiolino, R., Salvati, M., Risaliti, G., della Ceca,
  R., Matt, G., and Zamorani, G.: 2000,
\newblock {\em VizieR Online Data Catalog} {\bf 212}, 10473

\bibitem[\protect\astroncite{{Beckmann} et~al.}{2006}]{Beckmann_2006}
{Beckmann}, V., {Soldi}, S., {Shrader}, C.~R., {Gehrels}, N., and {Produit},
  N.: 2006,
\newblock {\em \apj} {\bf 652}, 126

\bibitem[\protect\astroncite{Bianchi et~al.}{2004}]{bianchi_x-ray_2004}
Bianchi, S., Matt, G., Balestra, I., Guainazzi, M., and Perola, G.~C.: 2004,
\newblock {\em Astronomy and Astrophysics} {\bf 422}, 65

\bibitem[\protect\astroncite{Boller et~al.}{2003}]{boller_xmm-newton_2003}
Boller, T., Keil, R., Hasinger, G., Costantini, E., Fujimoto, R., Anabuki, N.,
  Lehmann, I., and Gallo, L.: 2003,
\newblock {\em Astronomy and Astrophysics} {\bf 411}, 63

\bibitem[\protect\astroncite{{Brightman} and {Nandra}}{2008}]{Brightman_2008}
{Brightman}, M. and {Nandra}, K.: 2008,
\newblock {\em \mnras} {\bf 390}, 1241

\bibitem[\protect\astroncite{Cappi et~al.}{2006}]{cappi_x-ray_2006}
Cappi, M., Panessa, F., Bassani, L., Dadina, M., Dicocco, G., Comastri, A.,
  della Ceca, R., Filippenko, A.~V., Gianotti, F., Ho, L.~C., Malaguti, G.,
  Mulchaey, J.~S., Palumbo, G. G.~C., Piconcelli, E., Sargent, W. L.~W.,
  Stephen, J., Trifoglio, M., and Weaver, K.~A.: 2006,
\newblock {\em Astronomy and Astrophysics} {\bf 446}, 459

\bibitem[\protect\astroncite{Carrillo
  et~al.}{1999}]{carrillo_multifrequency_1999}
Carrillo, R., Masegosa, J., Dultzin-Hacyan, D., and Ordonez, R.: 1999,
\newblock {\em Revista Mexicana de Astronomia y Astrofisica} {\bf 35}, 187

\bibitem[\protect\astroncite{Ceca et~al.}{2002}]{ceca_enshrouded_2002}
Ceca, R.~D., Ballo, L., Tavecchio, F., Maraschi, L., Petrucci, P.~O., Bassani,
  L., Cappi, M., Dadina, M., Franceschini, A., Malaguti, G., Palumbo, G. G.~C.,
  and Persic, M.: 2002,
\newblock {\em Astrophysical Journal} {\bf 581}, L9

\bibitem[\protect\astroncite{{Dadina}}{2008}]{Dadina_2008}
{Dadina}, M.: 2008,
\newblock {\em \aap} {\bf 485}, 417

\bibitem[\protect\astroncite{{Della Ceca} et~al.}{2008}]{DellaCeca_2008}
{Della Ceca}, R., {Caccianiga}, A., {Severgnini}, P., {Maccacaro}, T.,
  {Brunner}, H., {Carrera}, F.~J., {Cocchia}, F., {Mateos}, S., {Page}, M.~J.,
  and {Tedds}, J.~A.: 2008,
\newblock {\em \aap} {\bf 487}, 119

\bibitem[\protect\astroncite{{Dewangan} and {Griffiths}}{2005}]{dewangan_2005}
{Dewangan}, G.~C. and {Griffiths}, R.~E.: 2005,
\newblock {\em \apjl} {\bf 625}, L31

\bibitem[\protect\astroncite{{Downes} and {Solomon}}{1998}]{downes+98}
{Downes}, D. and {Solomon}, P.~M.: 1998,
\newblock {\em \apj} {\bf 507}, 615

\bibitem[\protect\astroncite{Duc et~al.}{1997}]{duc_southern_1997}
Duc, P.-A., Mirabel, I.~F., and Maza, J.: 1997,
\newblock {\em Astronomy and Astrophysics Supplement Series} {\bf 124}, 533

\bibitem[\protect\astroncite{Dudik et~al.}{2005}]{dudik_chandra_2005}
Dudik, R.~P., Satyapal, S., Gliozzi, M., and Sambruna, R.~M.: 2005,
\newblock {\em Astrophysical Journal} {\bf 620}, 113

\bibitem[\protect\astroncite{{Dudik} et~al.}{2009}]{Dudik_2008}
{Dudik}, R.~P., {Satyapal}, S., and {Marcu}, D.: 2009,
\newblock {\em \apj} {\bf 691}, 1501

\bibitem[\protect\astroncite{{Evans} et~al.}{2002}]{evans+02}
{Evans}, A.~S., {Mazzarella}, J.~M., {Surace}, J.~A., and {Sanders}, D.~B.:
  2002,
\newblock {\em \apj} {\bf 580}, 749

\bibitem[\protect\astroncite{{Ferrarese} et~al.}{2001}]{Ferrarese_2001}
{Ferrarese}, L., {Pogge}, R.~W., {Peterson}, B.~M., {Merritt}, D., {Wandel},
  A., and {Joseph}, C.~L.: 2001,
\newblock {\em \apjl} {\bf 555}, L79

\bibitem[\protect\astroncite{Ghisellini
  et~al.}{1994}]{ghisellini_contribution_1994}
Ghisellini, G., Haardt, F., and Matt, G.: 1994,
\newblock {\em Monthly Notices of the Royal Astronomical Society} {\bf 267},
  743

\bibitem[\protect\astroncite{Gonzalez-Delgado
  et~al.}{2004}]{delgado_stellar_2004}
Gonzalez-Delgado, R.~M., Fernandes, R.~C., Perez, E., Martins, L.~P.,
  Storchi-Bergmann, T., Schmitt, H., Heckman, T., and Leitherer, C.: 2004,
\newblock {\em Astrophysical Journal} {\bf 605}, 127

\bibitem[\protect\astroncite{Gonzalez-Delgado
  et~al.}{2008a}]{gonzalez-delgado_hst/wfpc2_2008}
Gonzalez-Delgado, R.~M., Perez, E., Fernandes, R.~C., and Schmitt, H.: 2008a,
\newblock {\em Astronomical Journal} {\bf 135}, 747

\bibitem[\protect\astroncite{Gonzalez-Delgado
  et~al.}{2008b}]{gonzalez_delgado_hst/wfpc2_2008}
Gonzalez-Delgado, R.~M., Perez, E., Fernandes, R.~C., and Schmitt, H.: 2008b,
\newblock {\em Astronomical Journal} {\bf 135}, 747

\bibitem[\protect\astroncite{{Gonzalez-Martin}
  et~al.}{2009}]{gonzalez-martin_2009}
{Gonzalez-Martin}, O., {Masegosa}, J., {Marquez}, I., {Guainazzi}, M., and
  {Jimenez-Bailon}, E.: 2009,
\newblock {\em ArXiv e-prints}

\bibitem[\protect\astroncite{Gonzalez-Martin
  et~al.}{2006}]{gonzalez-martin_x-ray_2006}
Gonzalez-Martin, O., Masegosa, J., Marquez, I., Guerrero, M.~A., and
  Dultzin-Hacyan, D.: 2006,
\newblock {\em Astronomy and Astrophysics} {\bf 460}, 45

\bibitem[\protect\astroncite{{Gonzalez-Martin et
  al.}}{2008}]{gonzalez-martin_2008}
{Gonzalez-Martin et al.}, O.: 2008,
\newblock {\em PhD Thesis, Univerity of Granada, URL:
  http://www.star.le.ac.uk/~gmo4/O.Gonzalez-Martin-part1.pdf}

\bibitem[\protect\astroncite{Greenawalt et~al.}{1997}]{greenawalt_optical_1997}
Greenawalt, B., Walterbos, R. A.~M., and Braun, R.: 1997,
\newblock {\em Astrophysical Journal} {\bf 483}, 666

\bibitem[\protect\astroncite{Guainazzi et~al.}{2001}]{guainazzi_x-ray_2001}
Guainazzi, M., Fiore, F., Matt, G., and Perola, G.~C.: 2001,
\newblock {\em Monthly Notices of the Royal Astronomical Society} {\bf 327},
  323

\bibitem[\protect\astroncite{Guainazzi et~al.}{2005}]{guainazzi_x-ray_2005}
Guainazzi, M., Matt, G., and Perola, G.~C.: 2005,
\newblock {\em Astronomy and Astrophysics} {\bf 444}, 119

\bibitem[\protect\astroncite{{Ho}}{2008}]{ho_new_2008}
{Ho}, L.~C.: 2008,
\newblock {\em \araa} {\bf 46}, 475

\bibitem[\protect\astroncite{Ho et~al.}{2001}]{ho_detection_2001}
Ho, L.~C., Feigelson, E.~D., Townsley, L.~K., Sambruna, R.~M., Garmire, G.~P.,
  Brandt, W.~N., Filippenko, A.~V., Griffiths, R.~E., Ptak, A.~F., and Sargent,
  W. L.~W.: 2001,
\newblock {\em Astrophysical Journal} {\bf 549}, L51

\bibitem[\protect\astroncite{Ho et~al.}{1997}]{ho_search_1997}
Ho, L.~C., Filippenko, A.~V., Sargent, W. L.~W., and Peng, C.~Y.: 1997,
\newblock {\em Astrophysical Journal Supplement Series} {\bf 112}, 391

\bibitem[\protect\astroncite{Ibar and Lira}{2007}]{ibar_constraining_2007}
Ibar, E. and Lira, P.: 2007,
\newblock {\em Astronomy and Astrophysics} {\bf 466}, 531

\bibitem[\protect\astroncite{Imanishi et~al.}{2003}]{imanishi_x-ray_2003}
Imanishi, M., Terashima, Y., Anabuki, N., and Nakagawa, T.: 2003,
\newblock {\em Astrophysical Journal} {\bf 596}, L167

\bibitem[\protect\astroncite{Jimenez-Bailon
  et~al.}{2005}]{jimenez-bailon_xmm-newton_2005}
Jimenez-Bailon, E., Piconcelli, E., Guainazzi, M., Schartel, N.,
  Rodriguez-Pascual, P.~M., and Santos-Lleo, M.: 2005,
\newblock {\em Astronomy and Astrophysics} {\bf 435}, 449

\bibitem[\protect\astroncite{Keel}{1983}]{keel_spectroscopic_1983}
Keel, W.~C.: 1983,
\newblock {\em Astrophysical Journal} {\bf 269}, 466

\bibitem[\protect\astroncite{Keel et~al.}{1985}]{keel_eects_1985}
Keel, W.~C., Kennicutt, R.~C., Hummel, E., and van~der Hulst, J.~M.: 1985,
\newblock {\em Astronomical Journal} {\bf 90}, 708

\bibitem[\protect\astroncite{Koski}{1978}]{koski_spectrophotometry_1978}
Koski, A.~T.: 1978,
\newblock {\em Astrophysical Journal} {\bf 223}, 56

\bibitem[\protect\astroncite{{Krongold} et~al.}{2003}]{Krongold_2003}
{Krongold}, Y., {Dultzin-Hacyan}, D., and {Marziani}, P.: 2003,
\newblock in V. {Avila-Reese}, C. {Firmani}, C.~S. {Frenk}, and C. {Allen}
  (eds.), {\em Revista Mexicana de Astronomia y Astrofisica Conference Series},
  Vol.~17 of {\em Revista Mexicana de Astronomia y Astrofisica Conference
  Series}, pp 105--105

\bibitem[\protect\astroncite{{Lamastra} et~al.}{2009}]{Lamastra_2009}
{Lamastra}, A., {Bianchi}, S., {Matt}, G., {Perola}, G.~C., {Barcons}, X., and
  {Carrera}, F.~J.: 2009,
\newblock {\em ArXiv e-prints}

\bibitem[\protect\astroncite{{Leahy} and {Creighton}}{1993}]{Leahy_1993}
{Leahy}, D.~A. and {Creighton}, J.: 1993,
\newblock {\em \mnras} {\bf 263}, 314

\bibitem[\protect\astroncite{{Magdziarz} and
  {Zdziarski}}{1995}]{Magdziarz_1995}
{Magdziarz}, P. and {Zdziarski}, A.~A.: 1995,
\newblock {\em \mnras} {\bf 273}, 837

\bibitem[\protect\astroncite{Maiolino}{2001}]{maiolino_obscured_2001}
Maiolino, R.: 2001,
\newblock {\em X-ray Astronomy: Stellar Endpoints, AGN, and the Diffuse X-ray
  Background} {\bf 599}, 199

\bibitem[\protect\astroncite{Maiolino et~al.}{2003}]{maiolino_elusive_2003}
Maiolino, R., Comastri, A., Gilli, R., Nagar, N.~M., Bianchi, S., Beker, T.,
  Colbert, E., Krabbe, A., Marconi, A., Matt, G., and Salvati, M.: 2003,
\newblock {\em Monthly Notices of the Royal Astronomical Society} {\bf 344},
  L59

\bibitem[\protect\astroncite{Maiolino and
  Rieke}{1995}]{maiolino_low-luminosity_1995}
Maiolino, R. and Rieke, G.~H.: 1995,
\newblock {\em Astrophysical Journal} {\bf 454}, 95

\bibitem[\protect\astroncite{Maiolino et~al.}{1998}]{maiolino_heavy_1998}
Maiolino, R., Salvati, M., Bassani, L., Dadina, M., della Ceca, R., Matt, G.,
  Risaliti, G., and Zamorani, G.: 1998,
\newblock {\em Astronomy and Astrophysics} {\bf 338}, 781

\bibitem[\protect\astroncite{{Maoz}}{2007}]{Maoz_2007}
{Maoz}, D.: 2007,
\newblock {\em \mnras} {\bf 377}, 1696

\bibitem[\protect\astroncite{{Marziani} et~al.}{2003}]{Marziani_2003}
{Marziani}, P., {Sulentic}, J.~W., {Zamanov}, R., {Calvani}, M.,
  {Dultzin-Hacyan}, D., {Bachev}, R., and {Zwitter}, T.: 2003,
\newblock {\em \apjs} {\bf 145}, 199

\bibitem[\protect\astroncite{Matt}{1997}]{matt_x-ray_1997}
Matt, G.: 1997,
\newblock {\em Memorie della Societa Astronomica Italiana} {\bf 68}, 127

\bibitem[\protect\astroncite{{Merloni} et~al.}{2003}]{merloni_2003}
{Merloni}, A., {Heinz}, S., and {di Matteo}, T.: 2003,
\newblock {\em \mnras} {\bf 345}, 1057

\bibitem[\protect\astroncite{{Mihos} and {Hernquist}}{1994}]{mihos_h_1994}
{Mihos}, J.~C. and {Hernquist}, L.: 1994,
\newblock {\em \apjl} {\bf 437}, L47

\bibitem[\protect\astroncite{{Montero-Dorta} et~al.}{2008}]{Montero-Dorta_2008}
{Montero-Dorta}, A.~D., {Croton}, D.~J., {Yan}, R., {Cooper}, M.~C., {Newman},
  J.~A., {Georgakakis}, A., {Prada}, F., {Davis}, M., {Nandra}, K., and {Coil},
  A.: 2008,
\newblock {\em \mnras} pp 1355--+

\bibitem[\protect\astroncite{Moustakas and
  Kennicutt}{2006}]{moustakas_integrated_2006}
Moustakas, J. and Kennicutt, R.~C.: 2006,
\newblock {\em Astrophysical Journal} {\bf 651}, 155

\bibitem[\protect\astroncite{{Mulchaey} et~al.}{1994}]{Mulchaey_1994}
{Mulchaey}, J.~S., {Koratkar}, A., {Ward}, M.~J., {Wilson}, A.~S., {Whittle},
  M., {Antonucci}, R.~R.~J., {Kinney}, A.~L., and {Hurt}, T.: 1994,
\newblock {\em \apj} {\bf 436}, 586

\bibitem[\protect\astroncite{Nagar et~al.}{2005}]{nagar_radio_2005}
Nagar, N.~M., Falcke, H., and Wilson, A.~S.: 2005,
\newblock {\em Astronomy and Astrophysics} {\bf 435}, 521

\bibitem[\protect\astroncite{{Nandra} and {Iwasawa}}{2007}]{Nandra_2007}
{Nandra}, K. and {Iwasawa}, K.: 2007,
\newblock {\em \mnras} {\bf 382}, L1

\bibitem[\protect\astroncite{Nandra et~al.}{2007}]{nandra_xmm-newton_2007}
Nandra, K., O'Neill, P.~M., George, I.~M., and Reeves, J.~N.: 2007,
\newblock {\em Monthly Notices of the Royal Astronomical Society} {\bf 382},
  194

\bibitem[\protect\astroncite{{Nemmen} et~al.}{2006}]{Nemmen_2006}
{Nemmen}, R.~S., {Storchi-Bergmann}, T., {Yuan}, F., {Eracleous}, M.,
  {Terashima}, Y., and {Wilson}, A.~S.: 2006,
\newblock {\em \apj} {\bf 643}, 652

\bibitem[\protect\astroncite{Nenkova et~al.}{2002}]{nenkova_dust_2002}
Nenkova, M., Ivezic, Z., and Elitzur, M.: 2002,
\newblock {\em Astrophysical Journal} {\bf 570}, L9

\bibitem[\protect\astroncite{Osterbrock and
  Ferland}{2006}]{osterbrock_astrophysics_2006}
Osterbrock, D.~E. and Ferland, G.~J.: 2006,
\newblock {\em Astrophysics of gaseous nebulae and active galactic nuclei}

\bibitem[\protect\astroncite{Panessa and
  Bassani}{2002}]{panessa_unabsorbed_2002}
Panessa, F. and Bassani, L.: 2002,
\newblock {\em Astronomy and Astrophysics} {\bf 394}, 435

\bibitem[\protect\astroncite{Panessa et~al.}{2006}]{panessa_x-ray_2006}
Panessa, F., Bassani, L., Cappi, M., Dadina, M., Barcons, X., Carrera, F.~J.,
  Ho, L.~C., and Iwasawa, K.: 2006,
\newblock {\em Astronomy and Astrophysics} {\bf 455}, 173

\bibitem[\protect\astroncite{{Panessa} et~al.}{2008}]{panessa_broad-band_2008}
{Panessa}, F., {Bassani}, L., {de Rosa}, A., {Bird}, A.~J., {Dean}, A.~J.,
  {Fiocchi}, M., {Malizia}, A., {Molina}, M., {Ubertini}, P., and {Walter}, R.:
  2008,
\newblock {\em \aap} {\bf 483}, 151

\bibitem[\protect\astroncite{Panessa et~al.}{2005}]{panessa_nature_2005}
Panessa, F., Wolter, A., Pellegrini, S., Fruscione, A., Bassani, L., Ceca,
  R.~D., Palumbo, G. G.~C., and Trinchieri, G.: 2005,
\newblock {\em Astrophysical Journal} {\bf 631}, 707

\bibitem[\protect\astroncite{Perola et~al.}{2002}]{perola_compton_2002}
Perola, G.~C., Matt, G., Cappi, M., Fiore, F., Guainazzi, M., Maraschi, L.,
  Petrucci, P.~O., and Piro, L.: 2002,
\newblock {\em Astronomy and Astrophysics} {\bf 389}, 802

\bibitem[\protect\astroncite{{Peterson}}{1997}]{peterson_1997}
{Peterson}, B.~M.: 1997,
\newblock {\em {An Introduction to Active Galactic Nuclei}},
\newblock An introduction to active galactic nuclei, Publisher: Cambridge, New
  York Cambridge University Press, 1997 Physical description xvi, 238 p.~ISBN
  0521473489

\bibitem[\protect\astroncite{{Piconcelli} et~al.}{2005}]{piconcelli_2005}
{Piconcelli}, E., {Jimenez-Bail{\'o}n}, E., {Guainazzi}, M., {Schartel}, N.,
  {Rodr{\'{\i}}guez-Pascual}, P.~M., and {Santos-Lle{\'o}}, M.: 2005,
\newblock {\em \aap} {\bf 432}, 15

\bibitem[\protect\astroncite{Ptak et~al.}{2003}]{ptak_chandra_2003}
Ptak, A., Heckman, T., Levenson, N.~A., Weaver, K., and Strickland, D.: 2003,
\newblock {\em Astrophysical Journal} {\bf 592}, 782

\bibitem[\protect\astroncite{Rees}{1984}]{rees_black_1984}
Rees, M.~J.: 1984,
\newblock {\em Annual Review of Astronomy and Astrophysics} {\bf 22}, 471

\bibitem[\protect\astroncite{{Risaliti}}{2002}]{risaliti_2002b}
{Risaliti}, G.: 2002,
\newblock {\em \aap} {\bf 386}, 379

\bibitem[\protect\astroncite{Risaliti and
  Elvis}{2004}]{risaliti_panchromatic_2004}
Risaliti, G. and Elvis, M.: 2004,
\newblock Vol. 308, p. 187

\bibitem[\protect\astroncite{Risaliti
  et~al.}{1999}]{risaliti_distribution_1999}
Risaliti, G., Maiolino, R., and Salvati, M.: 1999,
\newblock {\em Astrophysical Journal} {\bf 522}, 157

\bibitem[\protect\astroncite{{Rovilos} et~al.}{2009}]{Rovilos_2009}
{Rovilos}, E., {Georgantopoulos}, I., {Tzanavaris}, P., {Pracy}, M., {Whiting},
  M., {Woods}, D., and {Goudis}, C.: 2009,
\newblock {\em ArXiv e-prints}

\bibitem[\protect\astroncite{{Rupke} et~al.}{2008}]{rupke+08}
{Rupke}, D.~S.~N., {Veilleux}, S., and {Baker}, A.~J.: 2008,
\newblock {\em \apj} {\bf 674}, 172

\bibitem[\protect\astroncite{{Sarzi} et~al.}{2005}]{Sarzi_2005}
{Sarzi}, M., {Rix}, H.-W., {Shields}, J.~C., {Ho}, L.~C., {Barth}, A.~J.,
  {Rudnick}, G., {Filippenko}, A.~V., and {Sargent}, W.~L.~W.: 2005,
\newblock {\em \apj} {\bf 628}, 169

\bibitem[\protect\astroncite{{Teng} et~al.}{2008}]{Teng_2008}
{Teng}, S.~H., {Veilleux}, S., {Anabuki}, N., {Dermer}, C.~D., {Gallo}, L.~C.,
  {Nakagawa}, T., {Reynolds}, C.~S., {Sanders}, D.~B., {Terashima}, Y., and
  {Wilson}, A.~S.: 2008,
\newblock {\em ArXiv e-prints}

\bibitem[\protect\astroncite{Terashima et~al.}{2002}]{terashima_x-ray_2002}
Terashima, Y., Iyomoto, N., Ho, L.~C., and Ptak, A.~F.: 2002,
\newblock {\em Astrophysical Journal Supplement Series} {\bf 139}, 1

\bibitem[\protect\astroncite{Tremaine et~al.}{2002}]{tremaine_slope_2002}
Tremaine, S., Gebhardt, K., Bender, R., Bower, G., Dressler, A., Faber, S.~M.,
  Filippenko, A.~V., Green, R., Grillmair, C., Ho, L.~C., Kormendy, J., Lauer,
  T.~R., Magorrian, J., Pinkney, J., and Richstone, D.: 2002,
\newblock {\em Astrophysical Journal} {\bf 574}, 740

\bibitem[\protect\astroncite{Turner et~al.}{1998}]{turner_asca_1998}
Turner, T.~J., George, I.~M., Nandra, K., and Mushotzky, R.~F.: 1998,
\newblock {\em Astrophysical Journal} {\bf 493}, 91

\bibitem[\protect\astroncite{Veilleux et~al.}{1995}]{veilleux_optical_1995}
Veilleux, S., Kim, D.-C., Sanders, D.~B., Mazzarella, J.~M., and Soifer, B.~T.:
  1995,
\newblock {\em Astrophysical Journal Supplement Series} {\bf 98}, 171

\bibitem[\protect\astroncite{{Veilleux} and
  {Osterbrock}}{1987}]{veilleux_high-resolution_1987}
{Veilleux}, S. and {Osterbrock}, D.~E.: 1987,
\newblock {\em \apjs} {\bf 63}, 295

\bibitem[\protect\astroncite{Veilleux et~al.}{1999}]{veilleux_new_1999}
Veilleux, S., Sanders, D.~B., and Kim, D.-C.: 1999,
\newblock {\em Astrophysical Journal} {\bf 522}, 139

\bibitem[\protect\astroncite{Vignali et~al.}{1999}]{vignali_probing_1999}
Vignali, C., Comastri, A., Cappi, M., Palumbo, G. G.~C., Matsuoka, M., and
  Kubo, H.: 1999,
\newblock {\em Astrophysical Journal} {\bf 516}, 582

\bibitem[\protect\astroncite{{Walsh} et~al.}{2008}]{Walsh_2008}
{Walsh}, J.~L., {Barth}, A.~J., {Ho}, L.~C., {Filippenko}, A.~V., {Rix}, H.-W.,
  {Shields}, J.~C., {Sarzi}, M., and {Sargent}, W.~L.~W.: 2008,
\newblock {\em \aj} {\bf 136}, 1677

\bibitem[\protect\astroncite{{Winter} et~al.}{2009}]{Winter_2009}
{Winter}, L.~M., {Mushotzky}, R.~F., {Reynolds}, C.~S., and {Tueller}, J.:
  2009,
\newblock {\em \apj} {\bf 690}, 1322

\bibitem[\protect\astroncite{{York} et~al.}{2006}]{York_2006}
{York}, D.~G., {Khare}, P., {Vanden Berk}, D., {Kulkarni}, V.~P., {Crotts},
  A.~P.~S., {Lauroesch}, J.~T., {Richards}, G.~T., {Schneider}, D.~P., {Welty},
  D.~E., {Alsayyad}, Y., {Kumar}, A., {Lundgren}, B., {Shanidze}, N., {Smith},
  T., {Vanlandingham}, J., {Baugher}, B., {Hall}, P.~B., {Jenkins}, E.~B.,
  {Menard}, B., {Rao}, S., {Tumlinson}, J., {Turnshek}, D., {Yip}, C.-W., and
  {Brinkmann}, J.: 2006,
\newblock {\em \mnras} {\bf 367}, 945

\bibitem[\protect\astroncite{{Zhang} et~al.}{2009}]{Zhang_2009}
{Zhang}, W.~M., {Soria}, R., {Zhang}, S.~N., {Swartz}, D.~A., and {Liu}, J.~F.:
  2009,
\newblock {\em \apj} {\bf 699}, 281

\end{thebibliography}

\scriptsize{
\begin{landscape}
\begin{longtable}{l c c c c c c c r c c c c}  
\caption{\normalsize{Properties of the LINER sample. }}  \\          
\hline\hline                
	  Name  		   &	 RA	&     Dec      & $\sf{N_{H1}}$  		&$\sf{N_{H2}}$  		  &${log~L_{X}}$ &  F([OIII])&  Av &HST       &  ${\sf M_{BH}}$	  &	  Ref($\sigma$) 	    &	Envir.	      \\
				   &		&	       &	&	  &    & &      &  &			&		      & 	   \\ 
(1)& 	      (2)		   &	 (3)	&      (4)     &  (5)				&  (6)  			  &   (7)  &(8) 	&(9)  &  (10)  &(11)     &  (12)       \\ \hline 
\endfirsthead
\caption{\normalsize{Continuation}}\\
\hline\hline                
	  Name  		   &	 RA	&     Dec      & $\sf{N_{H1}}$  		&$\sf{N_{H2}}$  		  &${log~L_{X}}$ &  F([OIII])&  Av &HST       &  ${\sf M_{BH}}$	  &	  Ref($\sigma$) 	    &	Envir.	      \\
				   &		&	       &	&	  &    & &      &  &			&		      & 	   \\ 
(1)& 	      (2)		   &	 (3)	&      (4)     &  (5)				&  (6)  			  &   (7)  &(8) 	&(9)  &  (10)  &(11)     &  (12)       \\ \hline 
\endhead
\endfoot
	   NGC\,0315\dotfill	   &00 57 48.88 &  +00 21 08.8 &  15.21 $\sf{_{10.67}^{ 21.33}}$&    1.06 $\sf{_{  0.87}^{  1.40}}$&   41.77  &     0.53&... & C&      8.61&  TD81    & 20    \\ 
	   NGC\,0410\dotfill	   &01 10 58.87 &  +03 09 08.3 &   0.12 $\sf{_{ 0.09}^{  0.17}}$&    0.01 $\sf{_{  0.01}^{  0.20}}$&   40.72  &     0.35&... &  &      8.79&  TD81    & 20    \\ 
	   NGC\,0474\dotfill	   &01 20 06.70 &  +00 24 55.0 &   ...  			&    ...			   &$<$40.26  &     0.34&0.61&  &      7.93&  TD81    & 2.5   \\ 
	 IIIZW\,035 \dotfill	   &01 44 30.50 &  +17 06 05.0 &   ...  			&    ...			   &   41.83  &    19.57&1.54& C&      ... & ...      & 1.5   \\ 
	   NGC\,0524\dotfill	   &01 24 47.72 &  +00 32 19.8 &   ...  			&    ...			   &   38.59  &     0.05&... & D&      8.29&  TD81    & 5     \\
	   NGC\,0833\dotfill	   &02 09 20.88 &  +10 08 00.3 &   3.51 $\sf{_{ 0.73}^{ 10.79}}$&   26.88 $\sf{_{  1.34}^{ 41.78}}$&   41.73  &     3.50&1.38&  &      8.97&  OFJSB   & 4     \\
	   NGC\,0835\dotfill	   &02 09 24.69 &$-$10 08 10.5 &   0.16 $\sf{_{ 0.01}^{  3.13}}$&   40.35 $\sf{_{ 24.60}^{ 69.46}}$&   41.40  &     ... &... &  &      8.16&  TD81    & 4     \\
	   NGC\,1052\dotfill	   &02 41 04.80 &  +08 15 20.8 &   0.36 $\sf{_{ 0.00}^{  1.88}}$& 12.90~$\sf{_{ 0.00}^{   35.47}}$ &   41.24  &    33.07&... & C&      8.07&  CDB-93  & 5     \\
	   NGC\,2639\dotfill	   &08 43 38.08 &  +50 12 20.0 &   0.80 $\sf{_{ 0.67}^{  0.92}}$&    ...			   &$<$40.06  &     4.74&0.78& C&      7.85&  S83     & 0     \\
	   NGC\,2655\dotfill	   &08 55 37.73 &  +78 13 23.1 &   0.01 $\sf{_{ 0.01}^{  0.07}}$&   30.20 $\sf{_{ 24.21}^{ 39.47}}$&   41.23  &    19.72&1.35& C&      7.10&  OFJSB   & 1.5   \\
	   NGC\,2681\dotfill	   &08 53 32.73 &  +51 18 49.3 &   0.15 $\sf{_{ 0.01}^{  0.28}}$&    0.01 $\sf{_{  0.01}^{  0.08}}$&   41.05  &     4.94&1.25& C&      6.70&  OFJSB   & 4     \\
	   NGC\,2685\dotfill	   &08 55 34.75 &  +58 44 03.9 &   ...  			&    ...			   &   40.82  &     5.28&1.56&  &      8.96&  Din+95  & 0?    \\
	   UGC\,4881\dotfill	   &09 15 55.10 &  +44 19 55.0 &   0.61 $\sf{_{ 0.32}^{  0.79}}$&    ...			   &$<$40.15  &     7.39&1.84& C&      ... & ...      & 1.5   \\
	    3C\,218 \dotfill	   &09 18 05.67 &$-$12 05 44.0 &   0.07 $\sf{_{ 0.04}^{  0.20}}$&    4.05 $\sf{_{  2.88}^{  5.96}}$&   42.08  &     0.38&... &  &      ... & ...      & 20    \\
	   NGC\,2787\dotfill	   &09 19 18.56 &  +69 12 12.0 &   0.11 $\sf{_{ 0.03}^{  0.22}}$&    ...			   &$<$38.81  &     1.17&... & C&      8.11&  TD81    & 0    \\
	   NGC\,2841\dotfill	   &09 22 02.63 &  +50 58 35.5 &   0.01 $\sf{_{ 0.01}^{  0.09}}$&    3.30 $\sf{_{  1.96}^{  5.60}}$&   39.22  &     1.41&0.22& C&      8.23&  WRF84   & 0    \\
	  UGC\,05101\dotfill	   &09 35 51.65 &  +61 21 11.3 &   0.21 $\sf{_{ 0.04}^{  0.66}}$&135.07 $\sf{_{43.51}^{  380.33}}$ &   43.84  &   561.20&4.67& C&      ... & ...      & 1.5   \\
	   NGC\,3185\dotfill	   &10 17 38.57 &  +21 41 17.7 &   ...  			&    ...			   &   41.15  &    24.89&1.33& C&      6.06&  NW95    & 4    \\
	   NGC\,3226\dotfill	   &10 23 27.01 &  +19 53 54.7 &   0.21 $\sf{_{ 0.11}^{  0.30}}$&    1.22 $\sf{_{  0.94}^{  1.76}}$&   40.80  &     1.33&... &  &      8.22&  TD81    & 1.5   \\
	   NGC\,3245\dotfill	   &10 27 18.39 &  +28 30 26.6 &   ...  			&    ...			   &   40.76  &     3.81&1.24& C&      8.39&  TD81    & 2.5   \\
	   NGC\,3379\dotfill	   &10 47 49.60 &  +12 34 53.9 &   ...  			&    ...			   &   39.91  &     0.68&... & C&      8.25&  TD81    & 4.5   \\
	   NGC\,3414\dotfill	   &10 51 16.23 &  +27 58 30.0 &   0.21 $\sf{_{ 0.13}^{  0.30}}$&    ...			   &   39.86  &     1.85&0.15&  &      8.67&  Din+95  & 5    \\
	   NGC\,3507\dotfill	   &11 03 25.39 &  +18 08 07.4 &   0.08 $\sf{_{ 0.01}^{  0.39}}$&    ...			   &$<$38.98  &     2.79&0.60& C&      ... & ...      & 2.5   \\
	   NGC\,3607\dotfill	   &11 16 54.66 &  +18 03 06.5 &   ...  			&    ...			   &   40.54  &     5.54&1.69& C&      8.48&  TD81    & 4.5   \\
	   NGC\,3608\dotfill	   &11 16 58.96 &  +18 08 54.9 &   ...  			&    ...			   &   39.98  &     0.28&... & C&      8.41&  TD81    & 5    \\
	   NGC\,3623\dotfill	   &11 18 55.96 &  +13 05 32.0 &   ...  			&    ...			   &$<$39.38  &     0.76&... & C&      7.62&  WKS     & 3    \\
	   NGC\,3627\dotfill	   &11 20 15.03 &  +12 59 29.6 &   ...  			&    ...			   &   41.19  &    27.92&1.87& D&      7.98&  WKS     & 3    \\
	   NGC\,3628\dotfill	   &11 20 17.01 &  +13 35 22.9 &   0.46 $\sf{_{ 0.41}^{  0.52}}$&    ...			   &   39.94  &     0.17&1.22& U&      7.91&  WKS     & 3    \\
	  NGC\,3690B\dotfill	   &11 28 32.20 &  +58 33 44.0 &   0.21 $\sf{_{ 0.03}^{  0.41}}$&    9.49 $\sf{_{  7.52}^{ 12.54}}$&   42.64  &   205.60&1.17& C&      ... & ...      & 1.5   \\
	   NGC\,3898\dotfill	   &11 49 15.37 &  +56 05 03.7 &   1.39 $\sf{_{ 1.12}^{  1.75}}$&    0.01 $\sf{_{  0.01}^{  0.59}}$&$<$40.55  &     0.90&... & C&      8.29&  S83     & 20    \\
	   NGC\,3945\dotfill	   &11 53 13.73 &  +60 40 32.0 &   0.04 $\sf{_{ 0.01}^{  0.17}}$&    ...			   &   39.12  &     0.55&0.27& C&      7.89&  OFJSB   & 0    \\
	   NGC\,3998\dotfill	   &11 57 56.12 &  +55 27 12.7 &   0.08 $\sf{_{ 0.06}^{  0.15}}$&    2.30 $\sf{_{  1.63}^{  3.18}}$&   41.32  &    32.16&1.22& C&      8.92&  TD81    & 5    \\
	   NGC\,4036\dotfill	   &12 01 26.75 &  +61 53 44.8 &   ...  			&    ...			   &   40.90  &     2.01&... & C&      8.12&  TD81    & 2.5   \\
	   NGC\,4111\dotfill	   &12 07 03.13 &  +43 03 55.4 &   4.67 $\sf{_{ 1.20}^{  9.65}}$&   37.71 $\sf{_{ 12.52}^{100.00}}$&$<$40.36  &     5.87&1.25& C&      7.56&  OFJSB   & 5    \\
	   NGC\,4125\dotfill	   &12 08 06.02 &  +65 10 26.9 &   0.53 $\sf{_{ 0.01}^{  0.88}}$&    0.86 $\sf{_{  0.08}^{  2.13}}$&$<$40.51  &     0.73&... & C&      8.31&  TD81    & 2    \\
IRAS\,12112+0305\dotfill	   &12 13 46.00 &  +02 48 38.0 &   ...  			&    ...			   &   43.00  &     2.36&1.11&  &      ... & ...      & 1    \\
	   NGC\,4261\dotfill	   &12 19 23.22 &  +05 49 30.8 &   0.69 $\sf{_{ 0.31}^{  1.38}}$&   16.45 $\sf{_{ 13.25}^{ 21.64}}$&   41.07  &     3.71&1.33& U&      8.96&  TD81    & 5    \\
	   NGC\,4278\dotfill	   &12 20 06.83 &  +29 16 50.7 &   0.09 $\sf{_{ 0.06}^{  0.12}}$&    2.65 $\sf{_{  1.48}^{  4.32}}$&   41.00  &     6.72&... & C&      8.46&  TD81    & 5    \\
	   NGC\,4314\dotfill	   &12 22 31.99 &  +29 53 43.3 &   0.27 $\sf{_{ 0.09}^{  0.44}}$&    0.01 $\sf{_{  0.01}^{  0.52}}$&$<$39.10  &     0.74&0.32& C&      7.19&  BHS02r  & 5    \\
	   NGC\,4321\dotfill	   &12 22 54.90 &  +15 49 20.6 &   0.59 $\sf{_{ 0.21}^{  0.77}}$&    0.19 $\sf{_{  0.08}^{  0.49}}$&   40.49  &     1.87&1.07& C&      6.76&  WK      & 20   \\
	   NGC\,4374\dotfill	   &12 25 03.74 &  +12 53 13.1 &   0.07 $\sf{_{ 0.01}^{  0.38}}$&    0.13 $\sf{_{  0.08}^{  0.25}}$&   41.31  &     3.45&1.19& C&      8.95&  TD81    & 10   \\
	  NGC\,4410A\dotfill	   &12 26 28.86 &  +09 01 10.8 &   0.51 $\sf{_{ 0.25}^{  1.14}}$&    0.01 $\sf{_{  0.01}^{  0.05}}$&   42.95  &     9.04&1.11& C&      ... & ...      & 4    \\
	   NGC\,4438\dotfill	   &12 27 45.59 &  +13 00 31.8 &   0.37 $\sf{_{ 0.26}^{  0.45}}$&    0.01 $\sf{_{  0.01}^{  0.21}}$&$<$40.83  &    15.56&1.75& D&      7.89&  TD81    & 2    \\
	   NGC\,4457\dotfill	   &12 28 59.01 &  +03 34 14.1 &   0.37 $\sf{_{ 0.08}^{  0.57}}$&    0.17 $\sf{_{  0.01}^{  2.60}}$&   40.59  &     4.69&0.57&  &      6.95&  OFJSB   & 0?    \\
	   NGC\,4459\dotfill	   &12 29 00.03 &  +13 58 42.8 &   ...  			&    ...			   &   38.37  &     0.23&0.13& C&      7.74&  TD81    & 5    \\
	   NGC\,4486\dotfill	   &12 30 49.42 &  +12 23 28.0 &   0.10 $\sf{_{ 0.09}^{  0.14}}$&    3.96 $\sf{_{  3.66}^{  4.47}}$&   40.82  &    10.80&0.94& C&      9.10&  TD81    & 10   \\
	   NGC\,4494\dotfill	   &12 31 24.03 &  +25 46 29.9 &   0.29 $\sf{_{ 0.01}^{  0.72}}$&    0.03 $\sf{_{  0.01}^{  0.17}}$&   38.78  &     0.20&... & C&      7.99&  TD81    & 0    \\
	   NGC\,4552\dotfill	   &12 35 39.81 &  +12 33 22.8 &   0.35 $\sf{_{ 0.01}^{  0.56}}$&    0.01 $\sf{_{  0.01}^{  0.12}}$&   39.25  &     0.37&0.09& C&      8.84&  TD81    & 0    \\
	   NGC\,4589\dotfill	   &12 37 25.03 &  +74 11 30.8 &   ...  			&    ...			   &   40.70  &     2.31&1.18& C&      8.57&  TD81    & 2    \\
	   NGC\,4579\dotfill	   &12 37 43.52 &  +11 49 05.5 &   0.48 $\sf{_{ 0.38}^{  0.54}}$&    0.45 $\sf{_{  0.27}^{  0.56}}$&   41.17  &     7.80&... & C&      8.09&  WKS     & 5    \\
	   NGC\,4596\dotfill	   &12 39 55.94 &  +10 10 33.9 &   ...  			&    ...			   &   38.47  &     0.12&... & C&      7.49&  TD81    & 2.5   \\
	   NGC\,4594\dotfill	   &12 39 59.43 &$-$11 37 23.0 &   0.19 $\sf{_{ 0.17}^{  0.23}}$&    ...			   &   39.97  &     6.30&0.24& C&      8.57&  Mar+94  & 0    \\
	   NGC\,4636\dotfill	   &12 42 49.87 &  +02 41 16.0 &   0.01 $\sf{_{ 0.01}^{  0.01}}$&    0.01 $\sf{_{  0.01}^{  0.17}}$&$<$39.03  &     0.25&... & D&      8.29&  TD81    & 0    \\
	  NGC\,4676A\dotfill	   &12 46 10.08 &  +30 43 55.2 &   ...  			&    ...			   &   39.85  &     0.23&1.58& D&      ... & ...      & 1.5   \\
	  NGC\,4676B\dotfill	   &12 46 11.23 &  +30 43 21.6 &   ...  			&    ...			   &   40.13  &     0.07&... & C&      8.79&  WHLD    & 1.5   \\
	   NGC\,4698\dotfill	   &12 48 22.92 &  +08 29 14.3 &   ...  			&    ...			   &   40.52  &     2.26&0.15& C&      7.87&  WK      & 0    \\
	   NGC\,4696\dotfill	   &12 48 49.28 &$-$41 18 40.0 &   0.01 $\sf{_{ 0.01}^{  0.18}}$&    0.01 $\sf{_{  0.01}^{  6.93}}$&   39.98  &     0.57&0.43& C&      8.56&  CDB-93  & 10   \\
	   NGC\,4736\dotfill	   &12 50 53.06 &  +41 07 13.6 &   0.31 $\sf{_{ 0.15}^{  0.61}}$&    0.04 $\sf{_{  0.02}^{  0.09}}$&   38.60  &     1.18&... & C&      7.43&  WKS     & 0    \\
	   NGC\,5005\dotfill	   &13 10 56.23 &  +37 03 33.1 &   0.61 $\sf{_{ 0.52}^{  0.80}}$&    0.01 $\sf{_{  0.01}^{  0.07}}$&$<$41.63  &     4.73&... & C&      ... &...       & 0     \\
	   NGC\,5055\dotfill	   &13 15 49.33 &  +42 01 45.4 &   0.16 $\sf{_{ 0.05}^{  0.33}}$&    ...			   &   39.56  &     3.05&1.62& C&      7.21&  FiBD86  & 4     \\
	  MRK\,266NE\dotfill	   &13 38 17.80 &  +48 16 41.2 &   0.01 $\sf{_{ 0.01}^{  0.28}}$&    9.45 $\sf{_{  5.69}^{ 28.94}}$&   43.43  &   180.90&0.62& C&      ... & ...      & 1     \\
	  UGC\,08696\dotfill	   &13 44 42.11 &  +55 53 12.7 &   0.60 $\sf{_{ 0.22}^{  0.96}}$&   50.91 $\sf{_{ 43.24}^{ 55.61}}$&   44.77  &   903.20&3.13& C&      7.74&  Jam+99  & 1     \\
   CGCG\,162-010\dotfill	   &13 48 52.43 &  +26 35 34.0 &   0.47 $\sf{_{ 0.39}^{  0.63}}$&    ...			   &   41.43  &     ... &... & C&      8.82&  OH      & 10    \\
	   NGC\,5363\dotfill	   &13 56 07.24 &  +05 15 17.0 &   0.01 $\sf{_{ 0.01}^{  0.08}}$&    2.66 $\sf{_{  1.74}^{  4.37}}$&   41.56  &    33.76&2.74&  &      8.12&  SCHL83  & 2     \\
	    IC\,4395\dotfill	   &14 17 21.08 &  +26 51 26.7 &   0.01 $\sf{_{ 0.01}^{  0.42}}$&    0.01 $\sf{_{  0.01}^{  0.17}}$&$<$42.58  &   320.80&6.05&  &      ... & ...      & 1.5   \\
IRAS\,14348-1447\dotfill	   &14 37 38.37 &$-$15 00 22.8 &   0.01 $\sf{_{ 0.01}^{  0.04}}$&    ...			   &   41.74  &     0.08&2.93& C&      ... &  ...     & 1     \\
	   NGC\,5746\dotfill	   &14 44 55.92 &  +01 57 18.0 &   0.60 $\sf{_{ 0.35}^{  0.93}}$&    ...			   &   40.22  &     1.35&2.24& C&      8.14&  BRBH93  & 2     \\
	   NGC\,5813\dotfill	   &15 01 11.26 &  +01 42 07.1 &   0.12 $\sf{_{ 0.01}^{  0.31}}$&    0.15 $\sf{_{  0.01}^{  0.51}}$&   40.55  &     0.19&... & C&      8.22&  TD81    & 5     \\
	   NGC\,5838\dotfill	   &15 05 26.26 &  +02 05 57.6 &   ...  			&    ...			   &   40.98  &     5.41&3.07& C&      8.75&  DS83    & 5     \\
	   NGC\,5846\dotfill	   &15 06 29.29 &  +01 36 20.2 &   0.28 $\sf{_{ 0.01}^{  0.43}}$&    0.03 $\sf{_{  0.01}^{  0.17}}$&$<$40.81  &     0.42&0.57& C&      8.49&  TD81    & 4.5   \\
	   NGC\,5866\dotfill	   &15 06 29.50 &  +55 45 47.6 &   ...  			&    ...			   &   40.07  &     1.75&2.29&  &      7.97&  TDT     & 5     \\
	   MRK\,0848\dotfill	   &15 18 06.35 &  +42 44 36.7 &   0.07 $\sf{_{ 0.01}^{  0.19}}$&    ...			   &   41.15  &     0.10&5.06& C&      ... & ...      & 1.5   \\
	   NGC\,6251\dotfill	   &16 32 31.97 &  +82 32 16.4 &   0.01 $\sf{_{ 0.01}^{  0.36}}$&    0.01 $\sf{_{  0.01}^{  0.04}}$&   43.36  &   127.00&4.59& C&      8.80&  H+85    & 2     \\
	   NGC\,6240\dotfill	   &16 52 58.89 &  +02 24 03.4 &   0.72 $\sf{_{ 0.55}^{  0.96}}$&   50.12 $\sf{_{ 22.34}^{103.79}}$&   44.19  &  2490.00&3.94& C&      8.84&  OOMM99  & 1     \\
IRAS\,17208-0014\dotfill	   &17 23 21.96 &  +00 17 00.9 &   0.34 $\sf{_{ 0.18}^{  0.61}}$&    ...			   &   42.97  &   137.60&6.39& C&      ... & ...      & 1     \\
	   NGC\,6482\dotfill	   &17 51 48.81 &  +23 04 19.0 &   0.19 $\sf{_{ 0.10}^{  0.33}}$&    ...			   &   41.11  &   524.30&... &  &      8.76&  7Sam    & 5     \\
	   NGC\,7130\dotfill	   &21 48 19.50 &$-$34 57 04.7 &   0.07 $\sf{_{ 0.04}^{  0.11}}$&   86.01 $\sf{_{ 60.51}^{160.48}}$&   42.57  &   159.20&1.30& C&      ... & ...      & 1.5?  \\
	   NGC\,7285\dotfill	   &22 28 38.00 &$-$24 50 26.8 &   0.68 $\sf{_{ 0.01}^{  0.78}}$&    0.87 $\sf{_{  0.59}^{  1.13}}$&   41.32  &     4.36&0.74&  &      ... & ...      & 1.5   \\
	   NGC\,7331\dotfill	   &22 37 04.09 &  +34 24 56.3 &   ...  			&    ...			   &   40.23  &     1.85&0.68& C&      7.56&  BRBH93  & 0     \\
	    IC\,1459\dotfill	   &22 57 10.60 &$-$36 27 44.0 &   0.20 $\sf{_{ 0.09}^{  0.28}}$&    1.26 $\sf{_{  0.66}^{  3.17}}$&   40.51  &     ... &... & C&      8.81&  CDB-93  & 4     \\
 NPM1G\,-12.0625\dotfill	   &23 25 19.82 &$-$12 07 26.4 &   0.71 $\sf{_{ 0.33}^{  0.85}}$&    0.15 $\sf{_{  0.06}^{  0.32}}$&$<$43.24  &     2.42&1.06& D&      8.18&  SHI90   & 10    \\
	   NGC\,7743\dotfill	   &23 44 21.14 &  +09 56 02.7 &   0.35 $\sf{_{ 0.18}^{  0.58}}$&    1.68 $\sf{_{  0.81}^{  3.63}}$&$<$41.33  &    46.37&1.84& C&      6.62&  Kor82   & 0    \\ 
\hline
\label{cha3:tab1}     
\end{longtable}
\noindent
Name (Col. 1), position (2000) (Cols. 2 and 3), 
soft (NH1, Col. 5) and hard (NH2, Col. 6) column densities in 
units of $\sf{10^{22}~cm^{-2}}$, logarith of the hard (2-10 keV) X-ray luminosity (Col. 7),
[OIII] emission line flux corrected for reddening in units of 
$\sf{1\times 10^{-14}erg~s^{-1}cm^{-2}}$.(Col. 8), optical extinction Av 
(Col. 9), {\it HST} morphology (Col. 10), black hole mass in logarithmical scale 
and their reference (Cols. 11 and 12) and environmental classification (Col. 13).
Av = 6.67 log(H$\sf{\alpha/Rv*H\beta}$). F([OIII]) measured by 
\citet{ho_detection_2001,moustakas_integrated_2006,veilleux_optical_1995,keel_eects_1985,
keel_spectroscopic_1983,koski_spectrophotometry_1978,greenawalt_optical_1997,
duc_southern_1997}.
Column (12) Interacting types: 
0 = Isolate,
1 = Merger,
1.5 = Close Interacting Pair,
2 = Pair,
2.5 = Wide Pair,
3 = Triplet,
4 = Compact Group,
4.5 = 1st group,
5 = Group,
10 = Cluster Center and,
20 = Cluster Member
\end{landscape}
}

\clearpage
\scriptsize{
\begin{longtable}{l c c c c c c c c c c}     
\caption{\normalsize{Summary of Compton-thick signatures. }}  \\      
\hline\hline        
	  Name  		   &${\sf \Gamma}$	&\emph{C-T1*} &${\sf F_{X}/F([OIII])}$  &\emph{C-T2*}&\multicolumn{3}{c}{EW(FeK$\alpha)$}&     \emph{C-T3*} & \emph{C-T}		  \\ \cline{6-8}
				   &                    &      &         &   &   Best-fit            & Pexrav			     &Baseline CT  &		     \\ 
(1)                                & 	      (2)	& (3)  & (4)     &(5)&(6)        &  (7)                  &(8)  				     &  (9)		 & (10)	 \\ \hline 
\endfirsthead
\caption{\normalsize{Continuation}}\\
\hline\hline
	  Name  		   &${\sf \Gamma}$	&\emph{C-T1*} &${\sf F_{X}/F([OIII])}$  &\emph{C-T2*}&\multicolumn{3}{c}{EW(FeK$\alpha)$}&     \emph{C-T3*} &\emph{C-T}		  \\ \cline{6-8}
				   &                    &      &         &   &   Best-fit            & Pexrav			     &Baseline CT  &		     \\ 
(1)                                & 	      (2)	& (3)  & (4)     &(5)&(6)        &  (7)                  &(8)  				     &  (9)		 & (10)	 \\ \hline 
\endhead
\endfoot
	   NGC\,0315\dotfill	   & 1.58 $\sf{_{ 1.42}^{ 2.08}}$ &      &   2.30  &   &      80 $\sf{_{    1}^{  160}}$& $<$  100			     & $<$     150				&		 &     \\ 
	   NGC\,0410\dotfill	   & $>$2.86			  &      &   1.38  &   & $<$  50  		      	& $<$  580			     & $<$     480				&		 &     \\ 
	   NGC\,0474\dotfill	   & ...			  &?     &  -0.13  &CT &     ...  		      	&        ...			     &  	  ...				&?		 &CT   \\ 
	 IIIZW\,035 \dotfill	   & ...			  &?     &  -1.34  &CT &     ...  		     	 &        ...			     &  	  ...				&?		 &CT   \\ 
	   NGC\,0524\dotfill	   & ...			&?     &   1.09  &   &     ...  		      	&        ...			     &  	  ...				&?		 &     \\
	   NGC\,0833\dotfill	   & ...			&?     &   1.78  &   &     330 $\sf{_{   50}^{  620}}$	&        ...			     & $<$     370			     $ \clubsuit$ & CT	 &CT?  \\
	   NGC\,0835\dotfill	   & ...			&?     &   ...   &?  &     770 $\sf{_{  480}^{ 1060}}$	&      630 $\sf{_{   380}^{   880}}$   &         610 $\sf{_{   360}^{    850}}$	&CT		 &CT   \\
	   NGC\,1052\dotfill	   & 1.27 $\sf{_{ 1.18}^{ 1.32}}$ &CT    &   1.07  &   &     140 $\sf{_{  120}^{  170}}$&       70 $\sf{_{    60}^{    90}}$   &         110 $\sf{_{    90}^{    130}}$	&		 &     \\
	   NGC\,2639\dotfill	   & ...		&?     &  -1.76  &CT &     ...  		      &        ...			     &  	  ...				&?		         &CT   \\
	   NGC\,2655\dotfill	   & 2.24 $\sf{_{ 1.88}^{ 4.26}}$ &      &   1.14  &   & $<$ 160  		      & $<$  140			     & $<$     110				&		 &     \\
	   NGC\,2681\dotfill	   & 1.51 $\sf{_{ 0.66}^{ 4.58}}$ &CT    &   0.08  &CT?& $<$2210  		      &        ...			     & $<$    1430				&CT		 &CT   \\
	   NGC\,2685\dotfill	   & ...		&?     &  -0.11  &CT &     ...  		      &        ...			     &  	  ...				&?		         &CT   \\
	   UGC\,4881\dotfill	   & ...		&?     &  -2.89  &CT &     ...  		      &        ...			     &  	  ...				&?		         &CT   \\
	    3C\,218 \dotfill	   & 2.60 $\sf{_{ 2.50}^{ 2.79}}$ &      &   1.75  &   & $<$  10  		      & $<$    3			     & $<$	 10				&		 &     \\
	   NGC\,2787\dotfill	   & 3.27 $\sf{_{ 2.02}^{ 5.16}}$ &      &   0.91  &   & $<$ 290  		      & $<$  250			     & $<$     190				&		 &     \\
	   NGC\,2841\dotfill	   & 1.95 $\sf{_{ 1.26}^{ 3.64}}$ &      &   0.84  &   & $<$ 240  		      & $<$  280			     & $<$     400				&		 &     \\
	  UGC\,05101\dotfill	   & 0.30 $\sf{_{-0.36}^{ 0.98}}$ &CT    &  -0.96  &CT &     280 $\sf{_{  100}^{  460}}$&      320 $\sf{_{   130}^{   500}}$   &         320 $\sf{_{   140}^{    510}}  \spadesuit$ &	 &CT   \\
	   NGC\,3185\dotfill	   & ...		&?     &  -0.70  &CT &     ...  		      &        ...			     &  	  ...				&?		         &CT   \\
	   NGC\,3226\dotfill	   & 1.81 $\sf{_{ 1.61}^{ 2.24}}$ &      &   1.85  &   & $<$ 110  		      & $<$   90			     & $<$     100				&		 &     \\
	   NGC\,3245\dotfill	   & ...		&?     &  -0.26  &CT &     ...  		      &        ...			     &  	  ...				&?		         &CT   \\
	   NGC\,3379\dotfill	   & ...		&?     &   0.17  &CT &     ...  		      &        ...			     &  	  ...				&?		         &CT   \\
	   NGC\,3414\dotfill	   & $>$2.51		&      &   0.72  &   & $<$ 590  		      & $<$ 7790			     & $<$     750				&CT		         &     \\
	   NGC\,3507\dotfill	   & ...		&?     &  -1.89  &CT &     ...  		      &        ...			     &  	  ...				&?		         &CT   \\
	   NGC\,3607\dotfill	   & ...		&?     &  -0.70  &CT &     ...  		      &        ...			     &  	  ...				&?		         &CT   \\
	   NGC\,3608\dotfill	   & ...		&?     &  -0.04  &CT &     ...  		      &        ...			     &  	  ...				&?		         &CT   \\
	   NGC\,3623\dotfill	   & ...		&?     &   1.70  &   &     ...  		      &        ...			     &  	  ...				&?		         &     \\
	   NGC\,3627\dotfill	   & ...		&?     &  -0.05  &CT &     ...  		      &        ...			     &  	  ...				&?		         &CT   \\
	   NGC\,3628\dotfill	   & 1.56 $\sf{_{ 1.38}^{ 2.10}}$ &      &   2.92  &   & $<$  80  		      & $<$   80			     & $<$	90				&		 &     \\
	  NGC\,3690B\dotfill	   & ...		&?     &  -0.72  &CT &     230 $\sf{_{  130}^{  340}}$&      220 $\sf{_{   110}^{   330}}$   &         250 $\sf{_{   140}^{    360}}$	&		         &CT?  \\
	   NGC\,3898\dotfill	   & ...		&?     &   0.06  &CT?&     ...  		      &        ...			     &  	  ...				&?		         &CT   \\
	   NGC\,3945\dotfill	   & ...		&?     &   0.61  &   & $<$ 110  		      &        ...			     & $<$	 2				&		         &     \\
	   NGC\,3998\dotfill	   & 1.88 $\sf{_{ 1.83}^{ 2.09}}$ &      &   1.48  &   & $<$  30  		      & $<$   20			     & $<$	40				&		 &     \\
	   NGC\,4036\dotfill	   & 2.14 $\sf{_{-0.97}^{ 6.54}}$ &CT    &  -0.03  &CT &     ...  		      &        ...			     &  	  ...				&?		 &CT   \\
	   NGC\,4111\dotfill	   & 1.37 $\sf{_{ 0.05}^{ 2.31}}$ &CT    &   1.22  &   & $<$ 180  		      & $<$  420			     & $<$     180				&		 &     \\
	   NGC\,4125\dotfill	   & $>$1.58		&      &   0.35  &CT?& $<$ 1560 		      & $<$  780			     & $<$    2380				&CT		         &CT   \\
IRAS\,12112+0305\dotfill	   & ...		&?     &  -0.11  &CT?&     ...  		      &        ...			     &  	  ...				&?		         &CT   \\
	   NGC\,4261\dotfill	   & 1.89 $\sf{_{ 1.54}^{ 2.40}}$ &      &   1.48  &   & $<$  30  		      & $<$   70			     & $<$	80				&		 &     \\
	   NGC\,4278\dotfill	   & 2.09 $\sf{_{ 1.93}^{ 2.40}}$ &      &  -0.09  &CT & $<$  50  		      & $<$   30			     & $<$	50				&		 &     \\
	   NGC\,4314\dotfill	   & 2.52 $\sf{_{ 0.06}^{ 4.66}}$ &CT    &   1.19  &   & $<$ 710  		      &       ...			     & $<$    1730				&CT		 &CT   \\
	   NGC\,4321\dotfill	   & 3.22 $\sf{_{ 1.64}^{ 6.25}}$ &      &   1.77  &   & $<$ 170  		      & $<$  240			     & $<$     240				&		 &     \\
	   NGC\,4374\dotfill	   & 2.59 $\sf{_{ 1.12}^{ 3.46}}$ &CT    &   0.44  &CT?& $<$1600  		      & $<$   50			     & $<$    1430				&CT		 &CT   \\
	  NGC\,4410A\dotfill	   & 1.46 $\sf{_{ 0.77}^{ 3.32}}$ &CT    &   0.19  &CT?& $<$ 430  		      & $<$  300			     & $<$     205				&		 &CT   \\
	   NGC\,4438\dotfill	   & 9.98 $\sf{_{ 0.86}^{ 9.95}}$ &CT    &  -0.60  &CT & $<$2460  		      &        ...			     & $<$    2020				&CT		 &CT   \\
	   NGC\,4457\dotfill	   & $>$0.79		&CT    &  -0.39  &CT & $<$  60  		      &        ...			     & $<$	60				&		         &CT   \\
	   NGC\,4459\dotfill	   & ...		&?     &   0.52  &   &      ... 		      &        ...			     &  	  ...				&?		         &     \\
	   NGC\,4486\dotfill	   & 2.79 $\sf{_{ 2.77}^{ 2.81}}$ &      &   1.34  &   &      90 $\sf{_{   40}^{  130}}$& $<$   50			     & $<$	50 $\sf{_{    10}^{    100}}$	&		 &     \\
	   NGC\,4494\dotfill	   & 1.91 $\sf{_{ 1.34}^{ 2.94}}$ &      &   0.93  &   & $<$ 370  		      &        ...			     & $<$     220				&		 &     \\
	   NGC\,4552\dotfill	   & 2.11 $\sf{_{ 1.82}^{ 3.58}}$ &      &   1.23  &   & $<$ 320  		      & $<$  420			     & $<$     390				&		 &     \\
	   NGC\,4589\dotfill	   & ...		&?     &  -0.16  &CT &      ... 		      &        ...			     &  	  ...				&?		         &CT   \\
	   NGC\,4579\dotfill	   & 1.71 $\sf{_{ 1.64}^{ 1.80}}$ &      &   1.75  &   &     110 $\sf{_{   70}^{  160}}$&       80 $\sf{_{    40}^{   120}}$   &         120 $\sf{_{    80}^{    160}}$	&		 &     \\
	   NGC\,4596\dotfill	   & ...		&?     &   0.84  &   &      ... 		      &        ...			     &  	  ...				&?		         &     \\
	   NGC\,4594\dotfill	   & 2.08 $\sf{_{ 1.76}^{ 2.36}}$ &      &   1.13  &   & $<$ 110  		      & $<$   80			     & $<$     100				&		 &     \\
	   NGC\,4636\dotfill	   & 1.63 $\sf{_{-0.54}^{ 4.91}}$ &CT    &   1.21  &   & $<$1750  		      & $<$ 2730			     & $<$    1890				&CT		 &CT   \\
	  NGC\,4676A\dotfill	   & ...		&?     &   0.58  &   &      ... 		      &        ...			     &  	  ...				&?		         &     \\
	  NGC\,4676B\dotfill	   & ...		&?     &   1.33  &   &      ... 		      &        ...			     &  	  ...				&?		         &     \\
	   NGC\,4698\dotfill	   & ...		&?     &  -0.13  &CT &      ... 		      &        ...			     &  	 ...				&?		         &CT   \\
	   NGC\,4696\dotfill	   & 3.65 $\sf{_{ 3.54}^{ 3.75}}$ &      &   1.06  &   &      ... 		      &        ...			     &  	  ...				&?		 &     \\
	   NGC\,4736\dotfill	   & 1.50 $\sf{_{ 1.39}^{ 1.79}}$ &      &   1.02  &   & $<$ 100  		      & $<$   40			     & $<$	50				&		 &     \\
	   NGC\,5005\dotfill	   & 2.03 $\sf{_{ 1.30}^{ 3.63}}$ &      &   0.45  &CT?&      ... 		      &        ...			     &  	 ...				&?		 &CT?  \\
	   NGC\,5055\dotfill	   & $>$-0.38		&CT    &  -0.42  &CT & $<$  40  		      &        ...			     & $<$	 1				&		         &CT   \\
	  MRK\,266NE\dotfill	   & ...		&?     &  -0.75  &CT &     280 $\sf{_{   70}^{  480}}$&      210 $\sf{_{    20}^{   390}}$   &         240 $\sf{_{    50}^{    440}}$	&		         &CT?  \\
	  UGC\,08696\dotfill	   & 0.33 $\sf{_{-0.42}^{ 0.85}}$ &CT    &   2.19  &   &     270 $\sf{_{  150}^{  380}}$&      240 $\sf{_{   120}^{   350}}$   &         250 $\sf{_{   130}^{    360}}$	&		 &     \\
   CGCG\,162-010\dotfill	   & 2.54 $\sf{_{ 2.45}^{ 2.60}}$ &      &   ...   &?  & $<$  10  		      & $<$    1			     & $<$	 5				&		 &     \\
	   NGC\,5363\dotfill	   & 1.51 $\sf{_{ 1.04}^{ 2.43}}$ &CT    &  -0.41  &CT & $<$ 340  		      & $<$  280			     & $<$     230				&		 &CT   \\
	    IC\,4395\dotfill	   & $>$1.13		&CT    &  -1.88  &CT &      ... 		      &        ...			     &  	 ...				&?		         &CT   \\
IRAS\,14348-1447\dotfill	   & $>$1.33		&      &   1.73  &   &      ... 		      &        ...			     &  	 ...				&?		         &     \\
	   NGC\,5746\dotfill	   & 1.51 $\sf{_{ 1.12}^{ 2.68}}$ &CT    &   1.17  &   & $<$ 380  		      & $<$  370			     & $<$     400				&		 &     \\
	   NGC\,5813\dotfill	   & $>$4.14		&      &   0.40  &CT?& $<$2610  		      & $<$ 8680			     & $<$    4880				&CT		         &CT   \\
	   NGC\,5838\dotfill	   &  ...		&?     &  -0.38  &CT &      ... 		      &        ...			     &  	 ...				&?		         &CT   \\
	   NGC\,5846\dotfill	   & 2.32 $\sf{_{ 1.71}^{ 3.18}}$ &      &   2.35  &   & $<$ 180  		      &        ...			     & $<$     200				&		 &     \\
	   NGC\,5866\dotfill	   & ...		&?     &  -0.30  &CT &      ... 		      &        ...			     &  	 ...				&?		         &CT   \\
	   MRK\,0848\dotfill	   & 1.47 $\sf{_{ 0.00}^{ 4.40}}$ &CT    &   1.66  &   & $<$2400  		      & $<$ 1020			     & $<$   30840				&CT		 &CT   \\
	   NGC\,6251\dotfill	   & 1.82 $\sf{_{ 1.77}^{ 1.89}}$ &      &  -0.33  &CT & $<$  50  		      & $<$   20			     & $<$	50				&		 &     \\
	   NGC\,6240\dotfill	   & 0.64 $\sf{_{ 0.52}^{ 0.93}}$ &CT    &  -0.88  &CT &     380 $\sf{_{  310}^{  440}}$&      410 $\sf{_{   350}^{   480}}$   &         460 $\sf{_{   390}^{    530}}  \spadesuit$ &	 &CT   \\
IRAS\,17208-0014\dotfill	   & 1.49 $\sf{_{-1.79}^{ 6.44}}$ &CT    &  -1.23  &CT & $<$ 560  		      & $<$ 6320			     & $<$    1330				&CT		 &CT   \\
	   NGC\,6482\dotfill	   & ...		&?     &  -2.90  &CT &      ... 		      &        ...			     &  	 ...				&?		         &CT   \\
	   NGC\,7130\dotfill	   & 0.12 $\sf{_{-0.86}^{ 0.49}}$ &CT    &  -1.05  &CT &     380 $\sf{_{  130}^{  630}}$&      480 $\sf{_{   200}^{   760}}$   &         420 $\sf{_{   160}^{    690}}$	&CT		 &CT   \\
	   NGC\,7285\dotfill	   & 1.82 $\sf{_{ 1.40}^{ 2.38}}$ &      &   1.12  &   &     210 $\sf{_{   40}^{  380}}$& $<$  300			     &         170 $\sf{_{    20}^{    320}}$	&		 &     \\
	   NGC\,7331\dotfill	   & ...		&?     &  -0.10  &CT &      ... 		      &        ...			     &  	 ...				&?		         &CT   \\
	    IC\,1459\dotfill	   & 2.63 $\sf{_{ 2.04}^{ 2.94}}$ &      &   ...   &?  & $<$  70  		      & $<$   80			     & $<$     120				&		 &     \\
 NPM1G\,-12.0625\dotfill	   & 3.00 $\sf{_{ 2.89}^{ 3.10}}$ &      &   0.01  &CT?& $<$  4			      & $<$    2			     & $<$	 3				&		 &     \\
	   NGC\,7743\dotfill	   & $>$-3.00		&CT    &  -0.75  &CT & $<$8340  		      &        ...			     & $<$    8240				&CT		         &CT   \\ 
\hline
\label{cha3:tab2}    
\end{longtable}
\noindent
Spectral 
Index (Col. 2), $\sf{F_{X}(2-10~keV)/F([OIII])}$ (Col. 4), EW(FeK$\alpha)$
for best-fit, pexrav model above 2 keV and the {\it baseline} model 
of {\it Compton-thick} AGN  by \citet{guainazzi_x-ray_2005} (Cols. 6, 7 and 8, respectively).
\emph{C-T1} = Compton-Thick candidates throught flat spectrum.
\emph{C-T2} = Compton-Thick candidates throught $\sf{F_{X}(2-10~keV)/F([OIII])}$ ratio.  
\emph{C-T3} = Compton-Thick candidates throught EW(FeKa).
\emph{C-T} = Final Compton-Thick candidates.
$\clubsuit$: NGC\,0833 is a {\it Compton-thin} object
with the {\it baseline} model of {\it Compton-thick} AGN.
$\spadesuit$: UGC\,05101 and NGC\,6240 are {\it Compton-thick} objects
with the {\it baseline} model of {\it Compton-thick} AGN.
}

\begin{table*}
\scriptsize{
  \caption{\scriptsize{$\sf{F_{X}(2-10~keV)/F([OIII])}$ ratios according to 
different environments.}}
 \begin{tabular}{lccr} \hline \hline \\
  Environment (Nr) & $\sf{F_{X}(2-10~keV)/F([OIII])}$         & $\sigma$ & range\\
          \\ \hline
ISO (15)  &  0.33 & 0.77 & -1.76 to 1.23\\
PAIR (29) & -0.06 & 1.02 & -2.89 to 2.19\\
GROUP(23) &  0.43 & 0.91 & -2.90 to 2.35\\
CLUSTER (9) & 1.12 & 1.12 & 0.01 to 2.30\\
\hline
\end{tabular}
\label{cha3:tab5}
}
\end{table*}
\end{document}